\documentclass[12pt]{article}
\title{Call center data analysis and model validation}
\author{Ger Koole$^{1,2}$, Siqiao Li$^2$, and Sihan Ding$^3$\\
\small $\strut^1$Vrije Universiteit Amsterdam, $\strut^2$CCmath bv, $\strut^3$Apple Inc}

\usepackage[dutch, english]{babel}
\usepackage[all]{xy}
\usepackage{amsmath, amssymb, amsthm, booktabs, rotating, multirow, grffile, pdfpages, setspace}
\usepackage{enumitem}
\usepackage{algorithm}
\usepackage[noend]{algpseudocode}
\usepackage{caption}
\usepackage{subcaption}
\usepackage[font=small, labelfont=bf, labelsep=period]{caption}

\newcommand{\p}{\mbox{P}}

\usepackage{graphicx}
\usepackage{bbm}
   \newcommand{\E}{\mathbbm{E}}
   
\usepackage{a4wide}
\usepackage{xcolor}

\begin{document}
\maketitle

\begin{abstract}
We analyze call center data on properties such as agent heterogeneity, customer patience and breaks. 
Then we compare simulation models that are different in the ways these properties are modeled.
We classify them according to the extend in which they approach the actual service level and average waiting times.
We obtain a theoretical understanding on how to distinguish between the model error and other aspects such as random noise.
We conclude that modeling explicitly breaks and agent heterogeneity is crucial for obtaining a precise model.
\end{abstract}

\section{Introduction}

Queueing and simulation models are used every day in tens of thousands of call centers worldwide to schedule many millions of call center agents.
However, very little attention has been paid as to whether these models give reliable unbiased estimates of the true performance.
One of the reasons is that, after the initial planning, many changes are made to the parameters of the model.
This can be driven by changes in the forecasted demand, but also by other factors such as the cancellation of meetings or agents being ill or requesting additional days off. 
However, even if changes are made to the agent schedules until the last moment, it is important to have precise models to make a good initial planning and make informed changes to that planning. 
Indeed, although models such as Erlang A model some crucial aspects of customer and agent behavior, it is not known whether all important aspects are modeled. 

In this paper we analyze first a data set which is unique in the sense that we do not only have the call logs with information about the calls, but also the agent logs with information about their breaks, the calls they handled, etc. 
This allows us for example to analyze agent heterogeneity and learning behavior.
See Gans et al.~\cite{GansLMSY} and Ibrahim et al.~\cite{IbrahimEST16-service-times} for other paper studying agent heterogeneity.
We also study aspects that are more commonly studied, such as the arrival process and abandonments.
See, e.g., Brown et al.~\cite{BrownGMSSZZ} for a seminal paper on call center data analysis, Ibrahim et al.~\cite{IbrahimYES-forecasting-survey} for a recent survey on call forecasting, and Whitt~\cite{Whitt-ms05-engineering} for an important paper on modeling abandonments.
To our knowledge breaks in cal centers have not yet been studied.

In Section \ref{results}, based on the data analysis in Section \ref{sec:analysis}, we construct and compare various models for this call center.
Because of some non-standard features, but also the fact that it concerns a multi-skill center, our analysis is based on simulations.
The models differ in the way the different aspects of the data analysis section are modeled. 
For example, do we take the average handling times constant over the whole year or do we make a model where they depend on the daily agent mix?
Another example of the way in which models differ is the model for breaks: do we follow the peaks of having paid short breaks found in the data or do we ignore them?
We validate the models by comparing the estimated performance of each model, using the simulations, to the realized performance. The comparison results show that ignoring certain features, such as agent breaks, agent heterogeneity and wrap-up times, will lead to highly inaccurate predictions of the performance measures.

When comparing the simulation results with the actual performance derived from the data we encounter some methodological issues.
Because the experimental conditions vary daily, we cannot simply compare the empirical distribution of achieved performance to that of the simulation model using standard statistical methods (as in, e.g., Sargent~\cite{Sargent}). 
The error we measure is the joint error of the model, the noise of the system, and the error caused by the (inevitably) incorrect forecast.
We develop a way to estimate the different errors and how to abstract the mean absolute error from the measured error.
This is done in Section~\ref{sec:set-up}.

We would like to finish this introduction by noting that many of the aspects discussed are typical for service operations in general, making this work also relevant to other fields than just call centers.

\section{Data analysis} 
\label{sec:analysis} 

Our data is from a multi-skill call center which is part of VANAD Laboratories in Rotterdam, for the year of 2014. 
The data consists of two separate data sets. 
One is the {\em call log data}, which gives the following information: customer call arrival times, departure times, the identity of the agent who handled the call, skill type of the call, etc. See Table \ref{table:calldata} for an example of this log. 
\begin{table}[htb]
\small
  \begin{center}
  \begin{tabular}{@{}ccccc@{}}
	\toprule
Call Arrival Time	& Skill ID	& Agent ID	& Answered time & Call Departure Time\\
   	\midrule
1/2/2014 8:03:21 & S & A	& 1/2/2014 08:33:22 &  1/2/2014 08:05:54\\
1/2/2014 8:04:37 & S & A	& 1/2/2014 08:04:38 &  1/2/2014 08:09:49\\
1/2/2014 8:06:36 & S & A	& 1/2/2014 08:06:27 &  1/2/2014 08:15:17\\
1/2/2014 8:08:07 & S & A	& 1/2/2014 08:08:08 &  1/2/2014 08:10:37\\
1/2/2014 8:08:26 & S & A	& 1/2/2014 08:08:27 &  1/2/2014 08:14:59\\
    \bottomrule
  \end{tabular}
  \caption{An example of the call data set}
  \label{table:calldata}
  \end{center}
\end{table}
The other one is referred to as the {\em activity data}, it describes agent activities. 
See Table \ref{table:actdata} for an example of this log. 

\begin{table}[htb]
\small
  \begin{center}
  \begin{tabular}{@{}cccc@{}}
	\toprule
Activity	& Start time	& End time	& Agent ID \\
   	\midrule
Logging in	& 1/2/2014 8:54:43 & 	1/2/2014 9:00:44	& A \\
Taking calls 	& 1/2/2014 9:00:44& 	1/2/2014 9:07:43& 	A\\
Meeting	& 1/2/2014 9:07:43	& 1/2/2014 9:08:47& 	A\\
Taking calls 	& 1/2/2014 9:08:47& 	1/2/2014 9:11:28& A \\
Wrap up	& 1/2/2014 9:11:28& 	1/2/2014 9:11:30	& A \\
    \bottomrule
  \end{tabular}
  \caption{An example of the activity data set}
  \label{table:actdata}
  \end{center}
\end{table}
The call log data has in total 1543164 call records from 27 different skills. 
There have been 312 agents working in this call center in the year 2014. 
This includes part-time agents, full-time agents, agents that worked only for a few months and agents that worked in every month of the year. 
Each agent has a skill set, which consists of at least one skill. Not every agent has all the skills. In this paper, we only focus on the top 8 skills that are most chosen by the customers, since they represent nearly 99\% of the total call volume. We explain how we deal with the amount of time that agents spend working on other skills while discussing the validation set up.

The routing mechanism is as follows. When a customer calls, she will first choose the call type from a menu. If there is any agent available with the skill to handle that type of calls, then she is routed to the longest idle agent of those available agents; otherwise, she will wait in a queue. The calls in this queue are served in the fashion of FCFS (first come first served), again according to agent skills. This is the common ``Longest Idle Agent - Longest Waiting Call" routing mechanism.

\subsection{Arrival moments}

\begin{figure}[ht]
        \centering
	  \begin{subfigure}[b]{0.48\textwidth}
                \includegraphics[width=\textwidth]{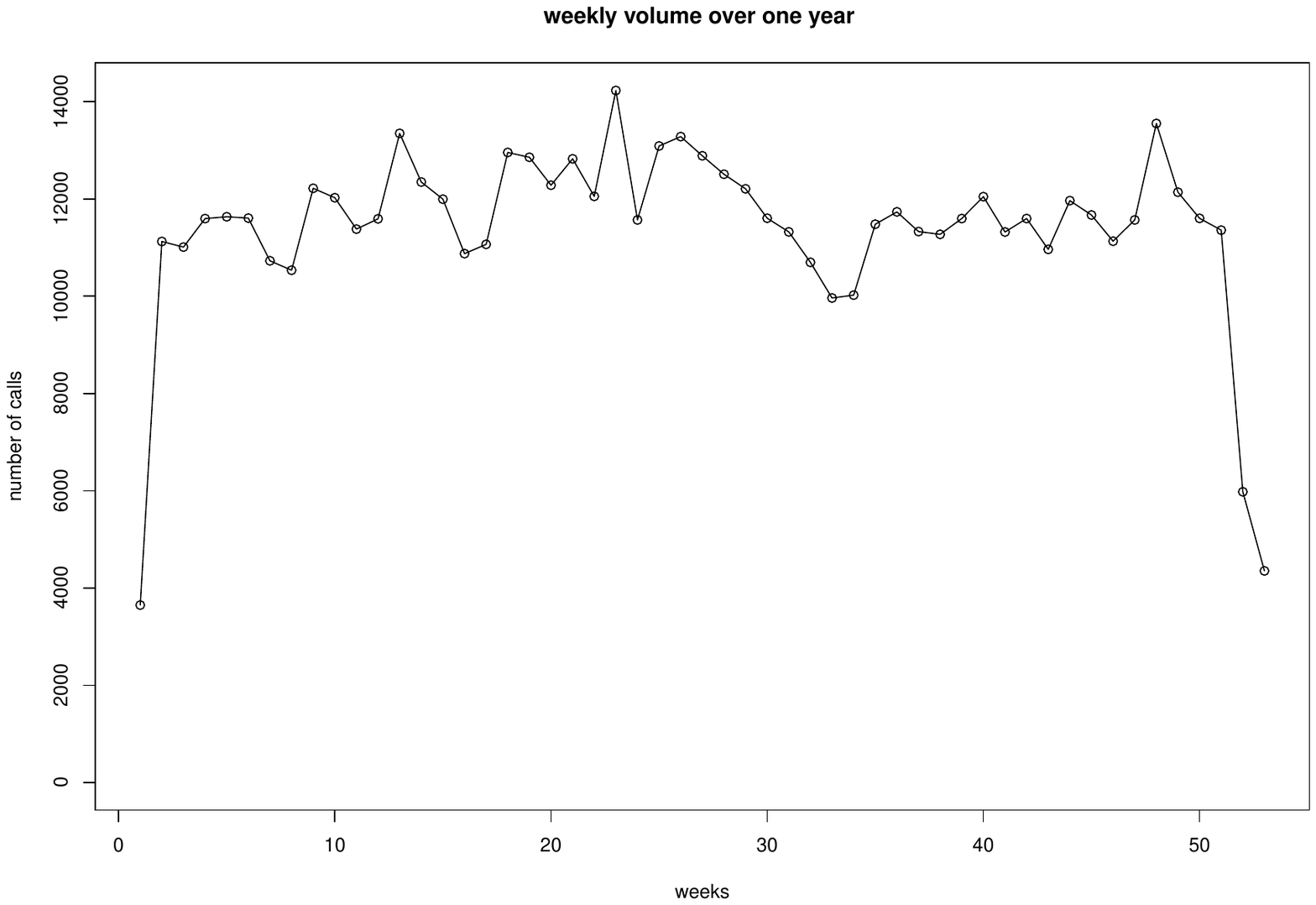}
                \label{fig:intra year}
        \end{subfigure}
        \begin{subfigure}[b]{0.48\textwidth}
                \includegraphics[width=\textwidth]{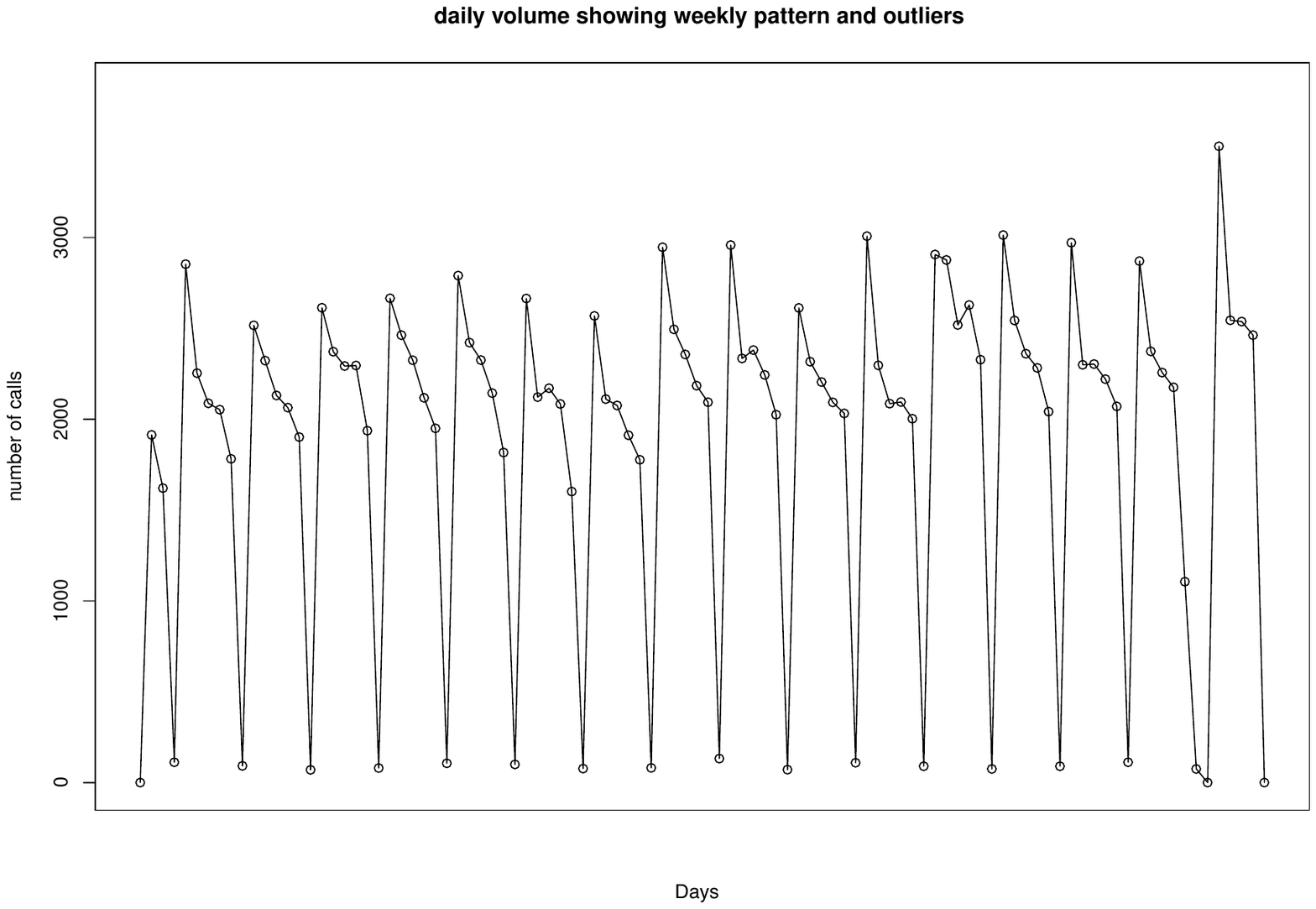}
                \label{fig:intra week}
        \end{subfigure} 
        \begin{subfigure}[b]{0.48\textwidth}
                \includegraphics[width=\textwidth]{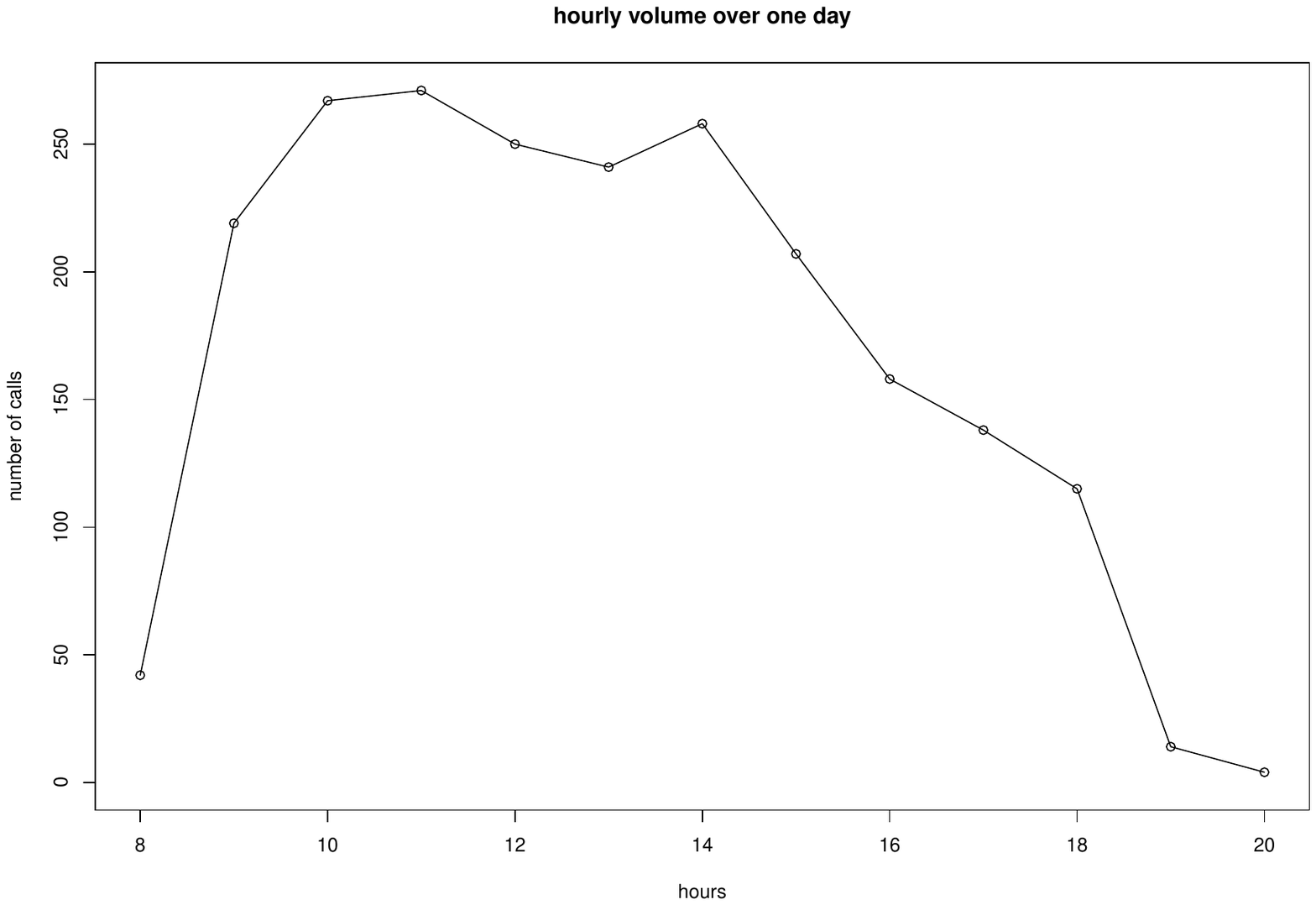}
                \label{fig:intra day}
        \end{subfigure} 
        \begin{subfigure}[b]{0.48\textwidth}
               \includegraphics[width=\textwidth]{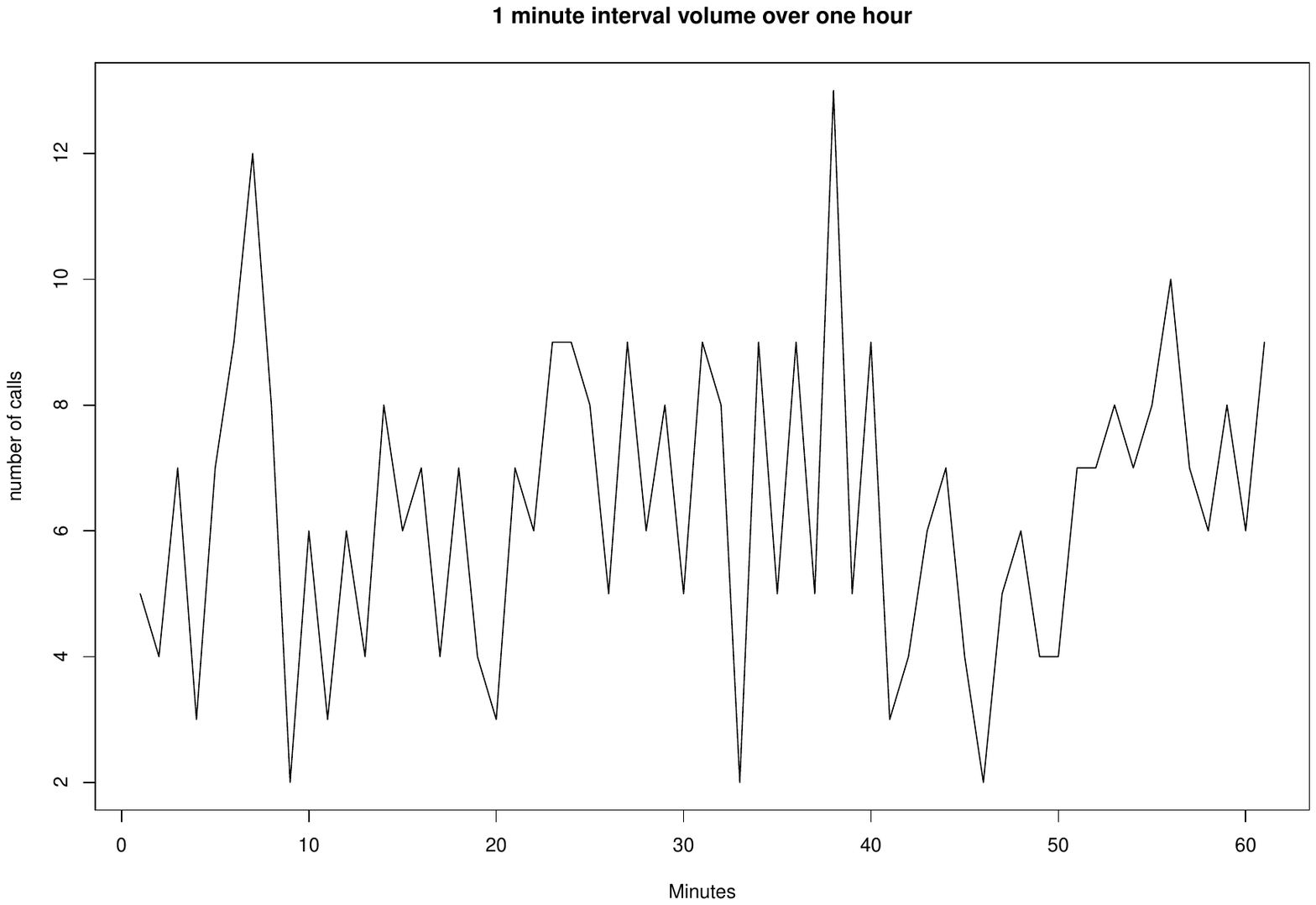}
                \label{fig:intra hour}
        \end{subfigure} 
       	 \caption{The non-homogeneous characteristics in arrival process}
		\label{fig:NHPP}
\end{figure}

Many papers have analysed call arrival processes for the sake of forecasting, see for example Ibrahim et al.~\cite{IbrahimYES-forecasting-survey} for a case study and references. 
They usually find seasonality in 3 time scales: intra-year, intra-week, and intra-day. At the lowest level there are random fluctuations. 
These four scales are also found in our data and plotted in Figure \ref{fig:NHPP} for skill ``30175". 

Our focus in this paper is not on the impact of forecasting errors, but on modeling assumptions.
One important assumption is that the arrivals form a nonhomogeneous Poisson process (NHPP) with intervals with piecewise constant (PC) rates, usually 15 or 30 minutes long.
This assumption is crucial for common performance models such as Erlang C and A and also usually made in simulation studies.  
In Section \ref{results} we quantify the impact of this assumption, combined with other assumptions resulting from the data analysis. 
However, it is of independent interest to test statistically whether the data is likely to come from an PC NHPP.

We refer to Kim \& Whitt \cite{kim2014call}  for background on testing for an NHPP. 
An obvious approach would be to test for exponential interarrival times. However, this requires the arrival rates to be known, which is not the case. 
An alternative characterization of the PP is that arrivals within an interval are uniformly distributed, and testing for this does not require assumptions on the arrival rate.
We tested for uniformity for intervals of different lengths (15, 30 and 60 minutes), for intervals separately or summed over whole days.
Our data is rounded to seconds, simulations shows that this has hardly any impact on $p$-values.
We did not use the Lewis transformation (as suggested in \cite{kim2014call}) because we wanted to keep the order of the arrivals to see if rates significantly change during intervals. 

In Table \ref{table:NHPPtest} we show some results of this Kolmogorov-Smirnov test.
We tested all intervals with a reduced confidence level of $1-\sqrt[k]{0.95}$ with $k$ the number of intervals with arrivals. 
For 15-minute intervals the null-hypothesis was nowhere rejected, for half hours in 1 interval, for hours in 2 intervals.
This supports the conclusion that call arrivals can well be modeled by a PC NHPP with 15-minute intervals.
Surprisingly, summed over all intervals the null-hypothesis was nowhere rejected.

\begin{table}[H]
  \begin{center}
  \begin{tabular}{@{}ccc@{}}
	\toprule
Time interval	& $N$	& $p$-value  \\
   	\midrule
8:00--8:15	& 11 & 	0.41\\
8:15--8:30 	& 13& 0.03	 \\
all quarters summed	& 2854      & 0.80   \\
8::00-8:30 	& 24  &  0.16   \\
8:30--9:00 	& 57& 	 $9\times10^{-3}$  \\
all half hours summed	& 2854      & 0.70   \\
8::00-9:00 	& 81  &  $1\times10^{-5}$    \\
9:00--10:00 	& 378& 	0.35 \\
all hours summed	& 2854	      & 0.782   \\
 	\bottomrule
  \end{tabular}
  \caption{NHPP tests for skill ``30175" on January 6, 2014}
  \label{table:NHPPtest}
 \end{center}
\end{table}

\subsection{Handling times}
\label{HT-sec}

Handling times (HT) in call centers have received less attention in the literature than arrival processes. 
However, as workload is the product of handling time and arrival forecast, it is equally important. 
Therefore, even small differences in the average handling time (AHT) can have a big impact on the SL estimation. 
Thus they are important to study. 
Gans et al.~\cite{GansLMSY} and Ibrahim et al.~\cite{IbrahimEST16-service-times} have shown that taking the time of day and which agents are working into account can result in more accurate predictions of the AHT. 
\cite{IbrahimEST16-service-times}~also conducted simple simulation experiments that demonstrated the impact of the AHT prediction model on the estimates of common performance measures in the call center. 
On the other hand, \cite{GansLMSY} focused on the learning effect of each agent and found different learning patterns. 

\begin{figure}[htb]
  \begin{center}
  \includegraphics[width=0.6\textwidth]{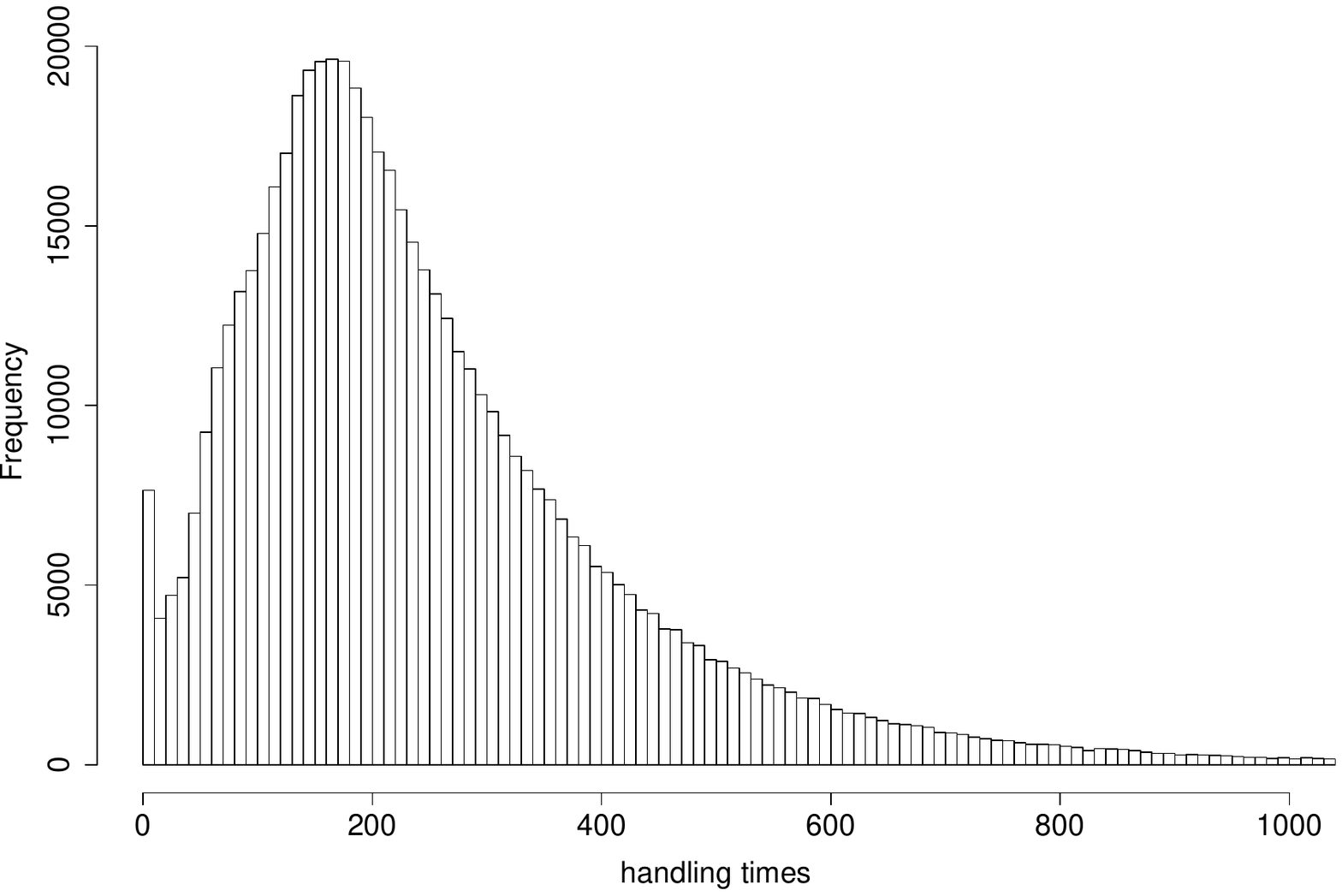}
  \caption{Handling times histogram of Skill 30175 (units: seconds)}
  \label{fig:HThist}
  \end{center}
\end{figure}

Figure~\ref{fig:HThist} displays the histogram of the handling times (HT) for Skill ``30175" in our dataset. 
We observe a peak around 0 seconds, which can be attributed to calls that end abruptly due to a loss of signal or call connection errors. It is worth noting that also Brown et al.~\cite{BrownGMSSZZ} reports a considerable number of short calls in their data, which is caused by agents who disconnect intentionally to have extra rest time. However, after consultation of the manager of VANAD Laboratories, we have determined that this is not the case in our call center. 

\begin{figure}[H]
        \centering
	  \begin{subfigure}[b]{0.48\textwidth}
                \includegraphics[width=\textwidth]{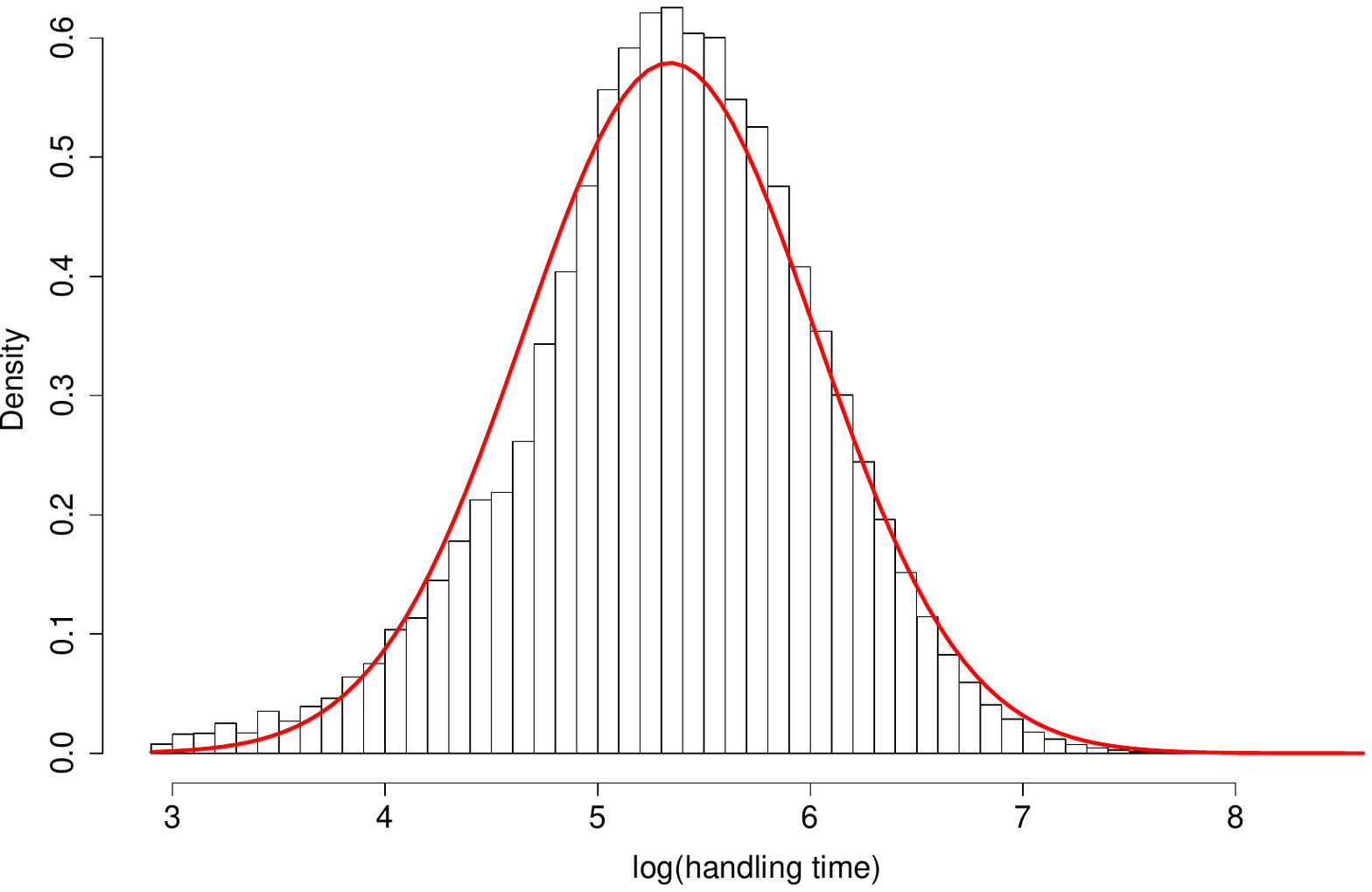}
                \label{fig:LHTfit}
        \end{subfigure}
        \begin{subfigure}[b]{0.48\textwidth}
                \includegraphics[width=\textwidth]{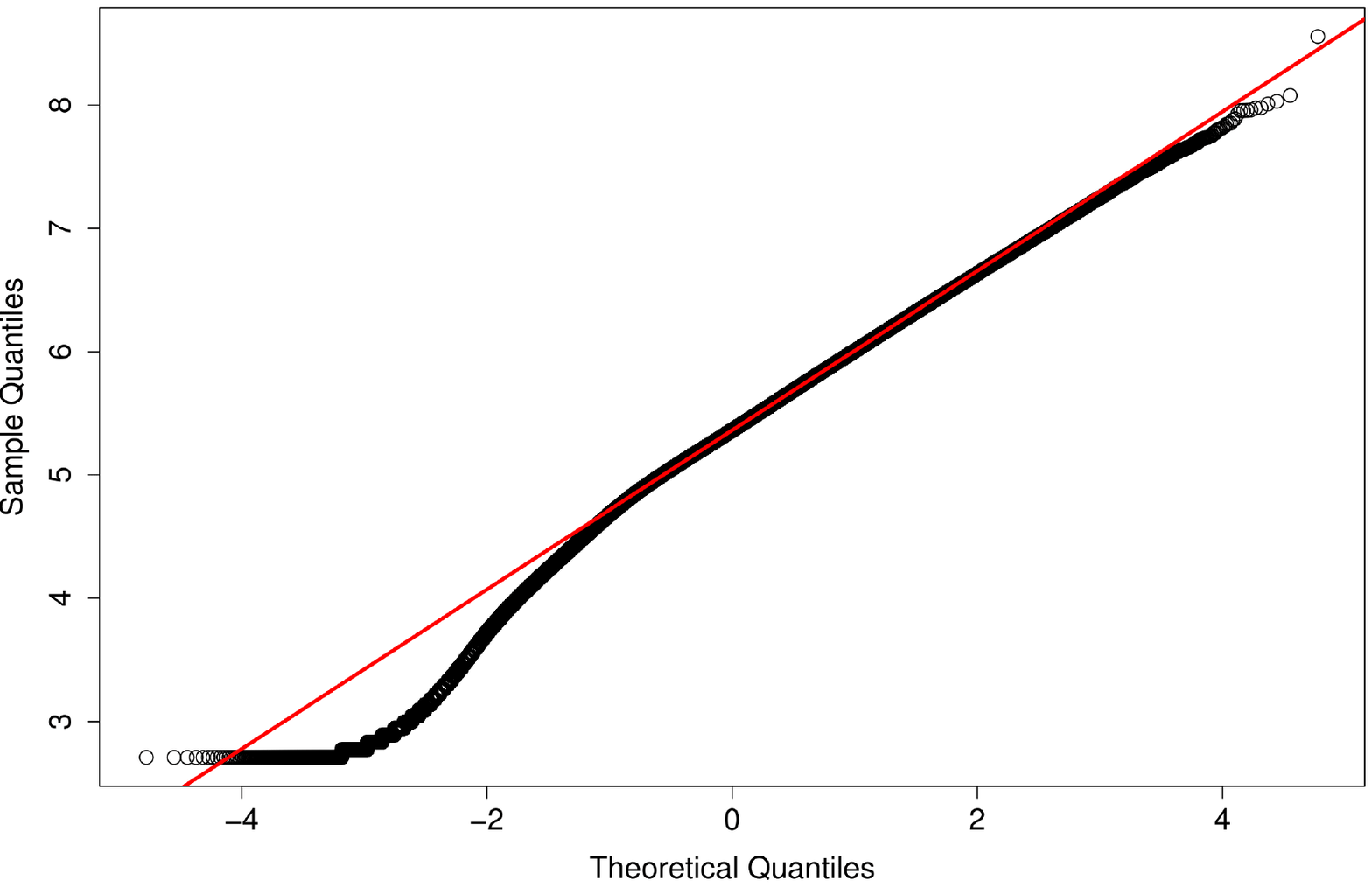}
                \label{fig:LHTqq}
        \end{subfigure} 
       	 \caption{Histogram (left) and Q-Q plot (right) of the log of the HT}
		\label{fig:logHT30175}
\end{figure}

We removed calls with HTs less than 15 seconds and fitted a log-normal distribution to the remaining data. 
Figure~\ref{fig:logHT30175} shows the histogram of the log with a normal density, as well as its Q-Q plot. 
We conclude that, similar to the findings of Bolotin~\cite{Bolotin} and Brown et al.~\cite{BrownGMSSZZ}, a log-normal distribution is a good fit for our HT data.

It is common to use the exponential distribution in models, although it is evidently not a good fit.
Assuming exponentially makes the analysis much simpler (e.g., the Erlang C formula for the M$\vert$M$\vert s$ queue versus the intractable M$\vert$G$\vert s$ queue), and multi-server systems are known to be relatively insensitive to higher moments of the handling times. 
The impact of assuming exponentially is studied numerically in Section \ref{results}.

\begin{figure}[htb]
  \begin{center}
    \includegraphics[width=0.6\textwidth]{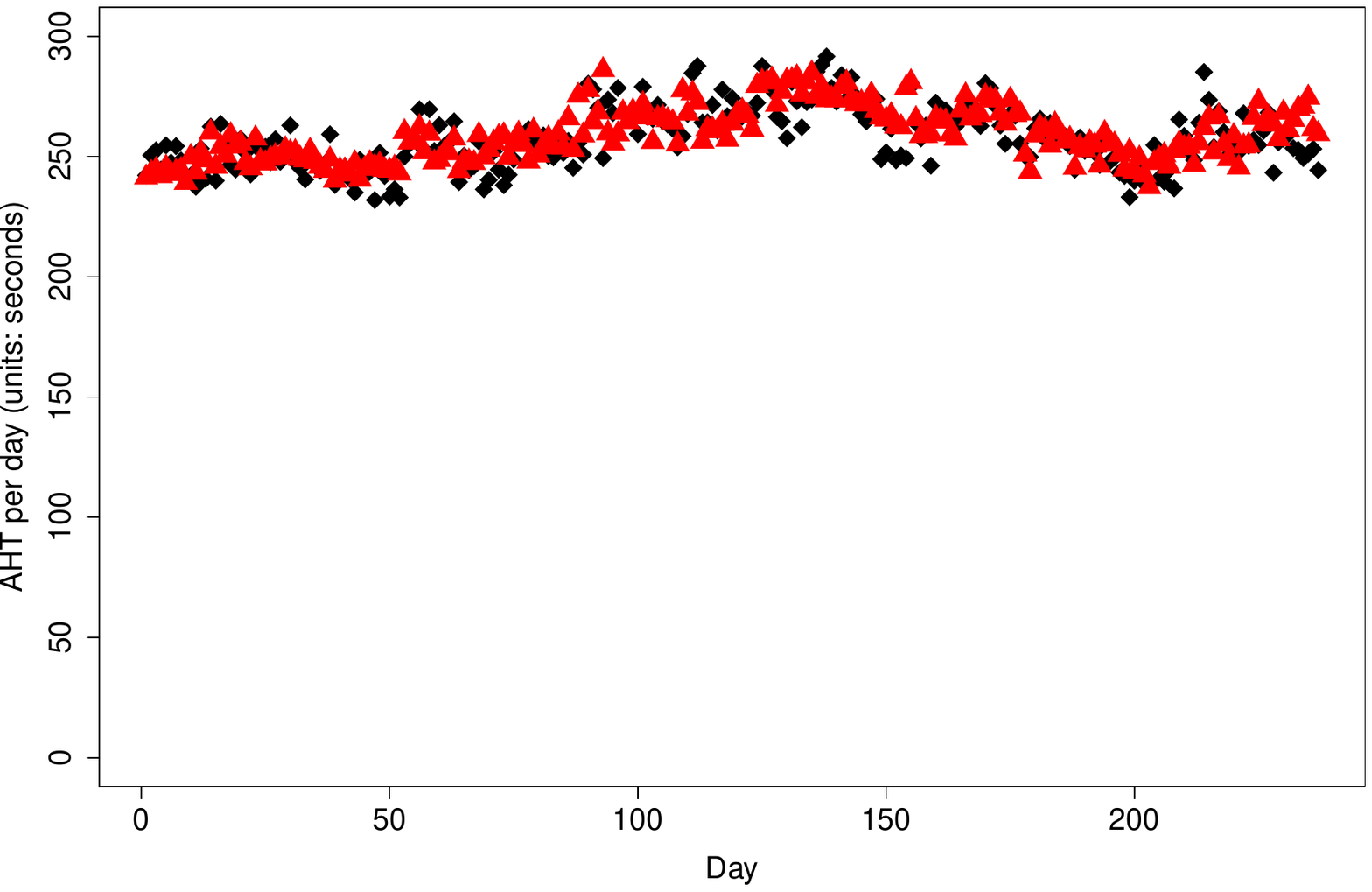}
  \caption{AHT per day, realisations (black diamonds) \& fit (red triangles)}
  \label{fig:AHTperdayfitted}
  \end{center}
\end{figure}

From now on we focus on the AHTs, and how it depends on other variables. 
Figure~\ref{fig:AHTperdayfitted} displays the AHT per day of Skill 30175, which demonstrate that the AHT varies from day to day. 
There can be three reasons for this variability:\\
- noise, although this can only explain part of the variability, given the number of calls per day;\\
- agent heterogeneity, some agents being, on average, faster than others;\\
- some external factor, for example a low SL that makes customers complain and take more time.

\begin{figure}[htb]
        \centering
	  \begin{subfigure}[b]{0.48\textwidth}
                \includegraphics[width=\textwidth]{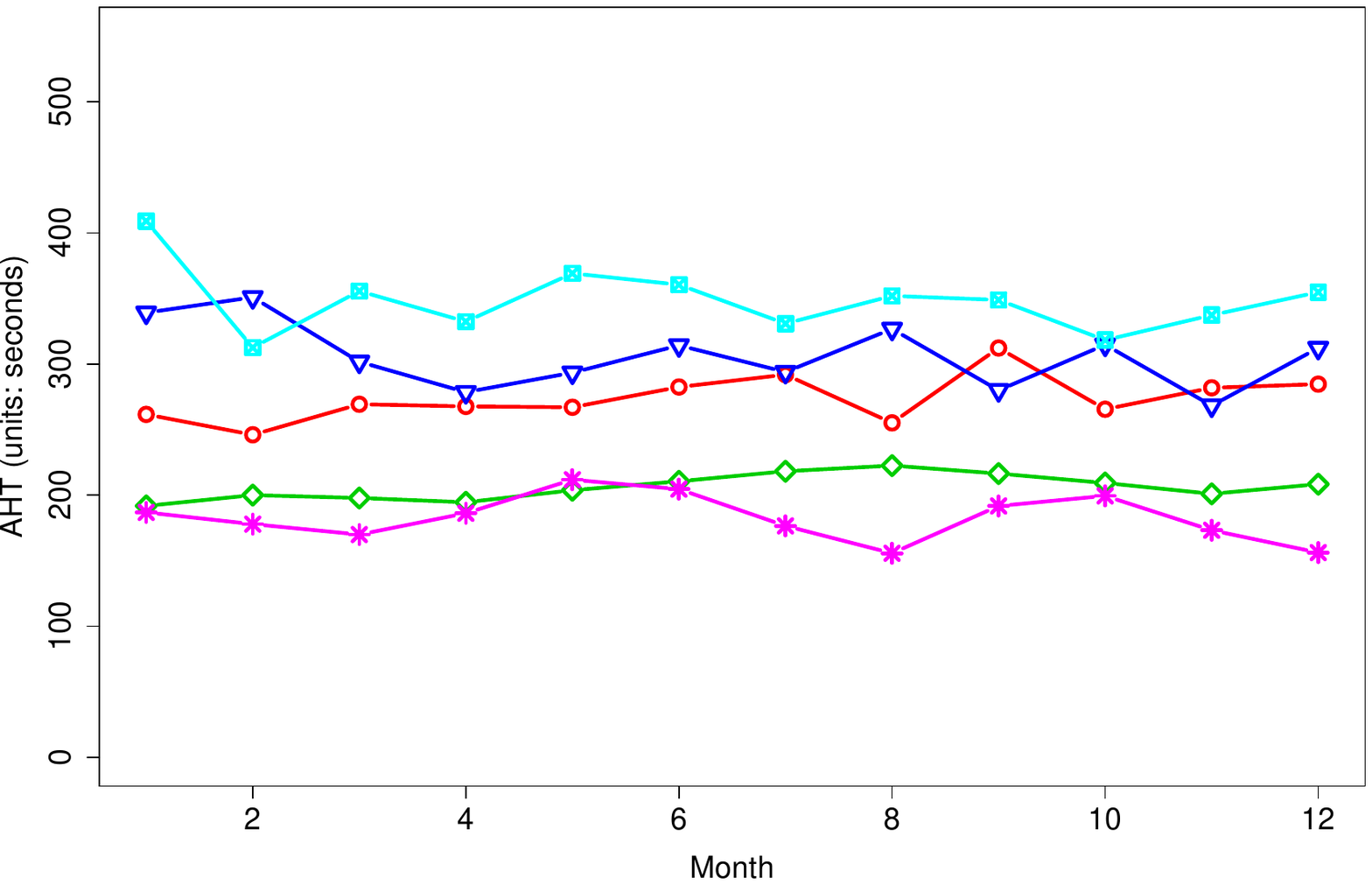}
                \label{fig:AHTexagents}
        \end{subfigure}
        ~ 
        \begin{subfigure}[b]{0.48\textwidth}
                \includegraphics[width=\textwidth]{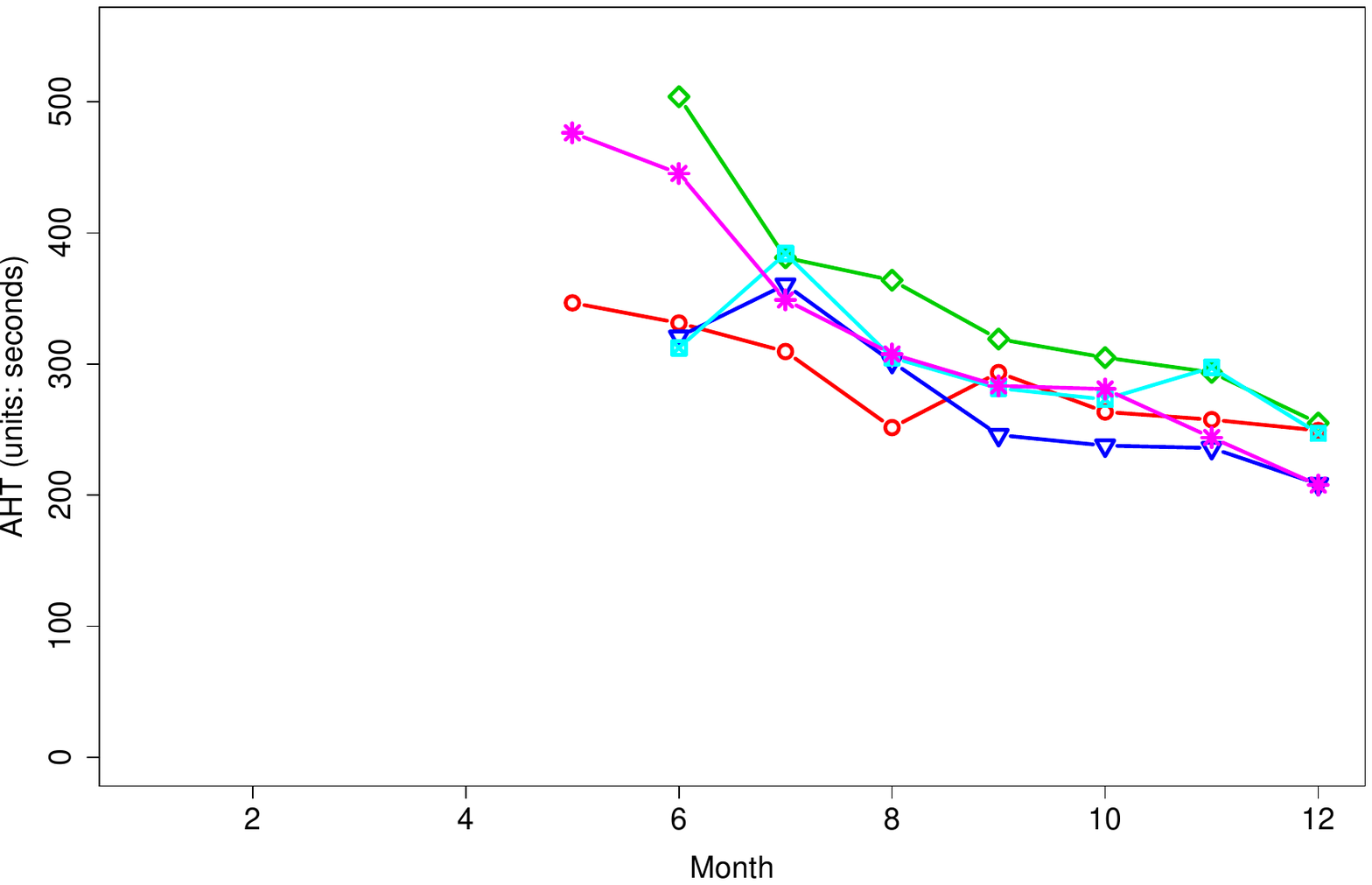}
                \label{fig:AHTnewagents}
        \end{subfigure} 
       	 \caption{AHT per month for experienced agents (left) and new agents (right)}
		\label{fig:AHTagents}
\end{figure}

To investigate the second reason in more depth, we plot the monthly AHTs of some experienced agents and some new agents in Figure~\ref{fig:AHTagents}. Note that we do not have the precise contract information of agents, hence we consider agents who have already started working since the beginning of the year as experienced agents, and the agents who only started in the middle of the year as new agents. 
We randomly select 5 agents from both groups and analyze their handling times.
As one can see from the left figure the AHT of each experienced agent is different. 
We conducted a Kolmogorov-Smirnov (KS) test which rejected the hypothesis that the daily AHTs of the 5 experienced agents come from the same distribution. 
The right figure shows that the AHTs of new agents have a decreasing trend, which suggests a learning effect. 
The KS test shows 3 agents are different from the other 2 agents in this group. 
We also conducted a KS test between the daily AHT data series of each pair of the agents, and 75.97\% of the tests rejected (p-value $\ll$ 0.05) the null hypothesis, which means the AHTs of agents do not come from the same distribution.

\begin{figure}[htb]
  \begin{center}
    \includegraphics[width=0.6\textwidth]{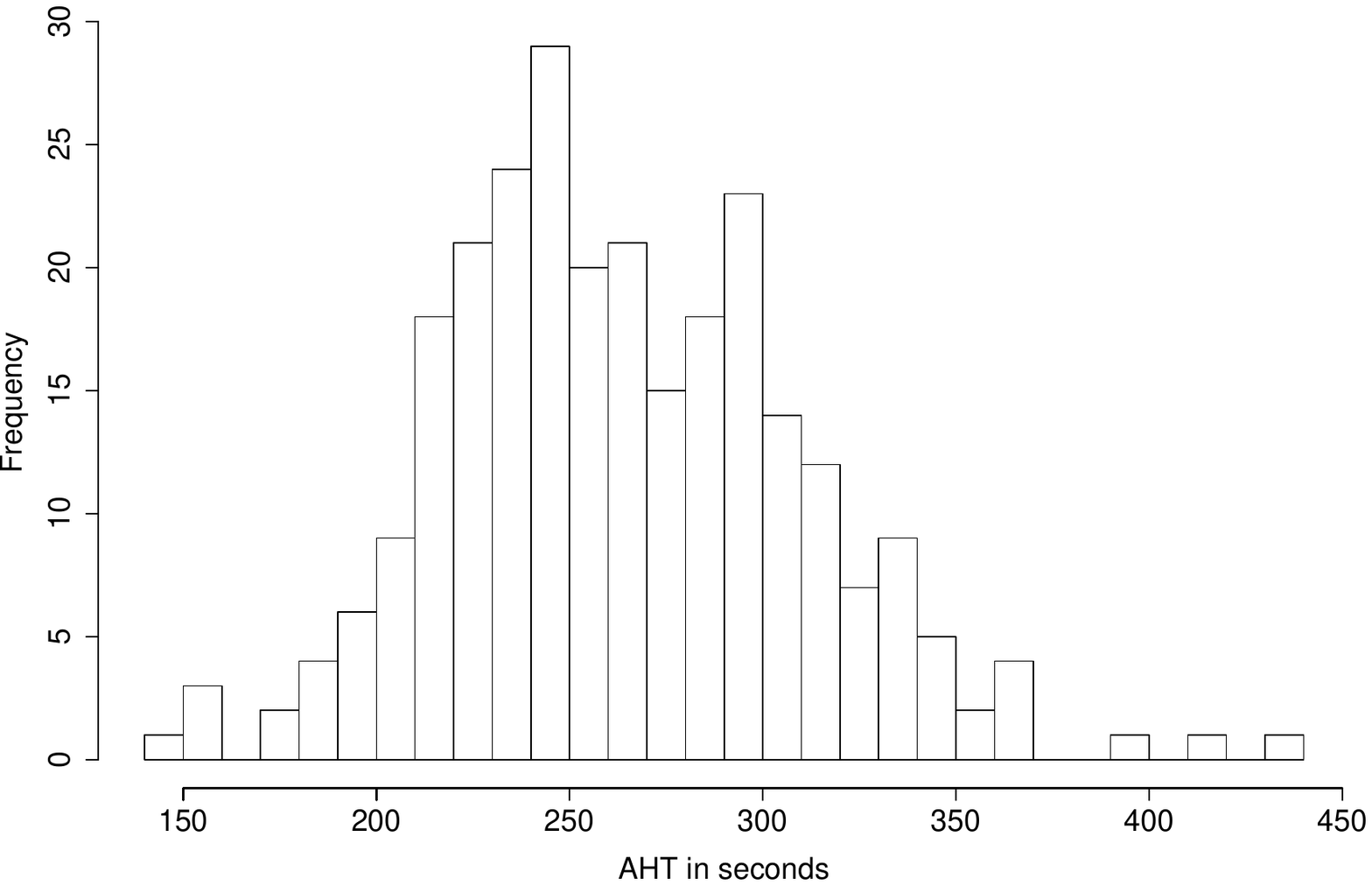}
  \caption{Histogram of the AHT }
  \label{fig:hisahtagent}
  \end{center}
\end{figure}

The histogram of the AHT of all agents for this specific skill is shown in Figure~\ref{fig:hisahtagent}, which further confirms the agent heterogeneity. 
To mitigate the effect of noise we only considered those agents who have answered more than 200 calls. 

We now develop a model that explicitly models learning and agent heterogeneity to predict the AHT for each agent for each day.
Let us first introduce some notation. 
Assume agent $j$ has handled $n_{jt}$ calls at day $t$, and call $i$ of agent $j$ at day $t$ had handling time $s_{jti}$.
We make a fit $\beta_{jt}$ for the AHT of agent $j$ at day $t$ with the following form:
\begin{align*}
\beta_{jt} = \alpha_j e^{-\gamma_jt},
\end{align*}
where $\alpha_j$ and $\gamma_j$ are such that 
\[\sum_t\sum_{i=1}^{n_{jt}}(\beta_{jt}-s_{jti})^2\]
is minimized.

Given the fit of AHT of each agent, we can then use the following model to fit the overall AHT $\beta_t$ of each day:
\begin{align}
\beta_t = {\sum_j n_{jt}\beta_{jt} \over \sum_j n_{jt}}. \label{eq:AHTfit}
\end{align}

The actual AHT of day $t$, $s_t=\sum_j\sum_ts_{jti}/\sum_jn_{jt}$, and the fitted AHT, $\beta_t$, are plotted in Figure~\ref{fig:AHTperdayfitted},  for all $t$. Figure~\ref{fig:AHTperdayfitted} suggests that the agent heterogeneity can explain a major part of the AHT variability. For each skill in Table~\ref{table:Rsquare} we calculate the explained variation $R^2$, which is defined by 
\begin{align*}
R^2 := 1 - {\sum_t(s_t-\beta_t)^2\over\sum_t(s_t-\bar s)^2},
\end{align*}
where $\bar s$ is the overall AHT, i.e., $\bar s=\sum_j\sum_t\sum_is_{jti}/\sum_j\sum_tn_{jt}$.

\begin{table}[H]
  \begin{center}
  \begin{tabular}{@{}ccccccccc@{}}
	\toprule
 Skill & 30175  & 30560  & 30172  & 30181 & 30179 & 30066 & 30518 & 30214\\
   	\midrule
 $R^2$ & 0.636 & 0.600 & 0.558 & 0.520 & 0.324 & 0.145 & 0.515 & 0.317\\
    \bottomrule
  \end{tabular}
  \caption{$R^2$ values of $\widehat{\text{AHT}}_t$.}
  \label{table:Rsquare}
  \end{center}
\end{table}
As one can see from Table~\ref{table:Rsquare} the $R^2$ for the first four skills, including skill 30185, is higher than 50\%. 
Thus the AHT prediction model in Equation~\eqref{eq:AHTfit} explains more than half of the variability in AHT. 
The explained variability is lower for the remaining 4 skills.
The small sample sizes of these skills can explain this.

In most call centers, the agent's workload does not end the moment the call ends, since agents often still have to do some {\em after-call work}, also known as {\em wrap-up time}. 
We have not yet found any empirical or theoretical results on the wrap-up times in the literature. 
We think this is mainly because wrap-up times are difficult to measure and they are usually not recorded in call center data. In our data set, the wrap-up has a specific activity ID (i.e., ID = 16) and its start and end times are recorded in the agent activity data file. This allows us to empirically study the wrap-up times and discuss the impact of ignoring them in the workforce management models.

\begin{figure}[htb]
  \begin{center}
    \includegraphics[width=0.6\textwidth]{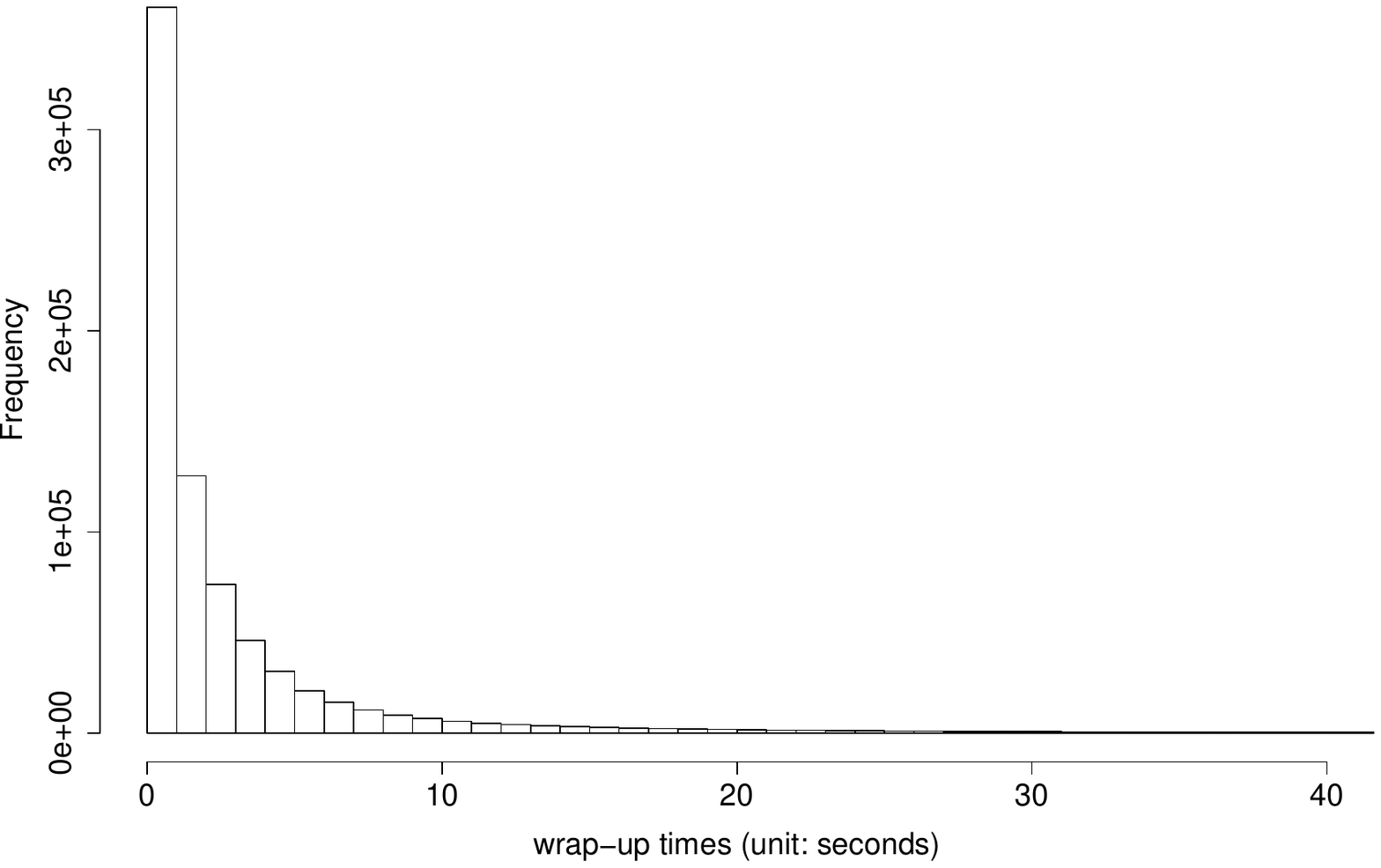}
  \caption{Wrap-up times histogram}
  \label{fig:WThist}
  \end{center}
\end{figure}

In Figure~\ref{fig:WThist}, we plot the histogram of the wrap-up times. As one can see the wrap-up times in this data set are in general quite short and often have a duration of 0 second. The average wrap-up time is approximately 3.28 seconds. It would be even better if the wrap-up time is recorded per call in the call log file. In that way, we can plot the wrap-up time histogram per skill, which may follow different distributions, especially in large call centers.  
Although the wrap-up time is short, it should be considered as part of the handling time because while wrapping up an agent is still occupied and cannot handle the next call. We also noticed that some call centers may have a threshold (e.g., 60 seconds) regarding the length of wrap-up times to prevent agents from using the time to take a break. However, this often results in a high peak at the threshold length.

\subsection{Patience}

Customers' impatience is known to have a drastic effect on system performance (Gans et al.~\cite{GansKM}). 
It is directly related to an important performance metric: the abandonment rate (some call centers prefer to use connected rate which is 1 $-$ abandonment rate). 
The patience is defined as the time that a caller is willing to wait in the queue before getting service.
Thus, callers leave the queue (i.e., abandon) if they are not handled before their patience is finished.
Vice versa, if the call gets answered before the patience is over, the patience of this caller is truncated at the waiting time. 
Therefore, the records of the time spent in the queue is a right censored data set for patience. 
To statistically estimate the patience distribution, we can use the famous Kaplan-Meier estimator (\cite{KaplanMeier}).

\begin{figure}[htb]
        \centering
	  \begin{subfigure}[b]{0.48\textwidth}
                \includegraphics[width=\textwidth]{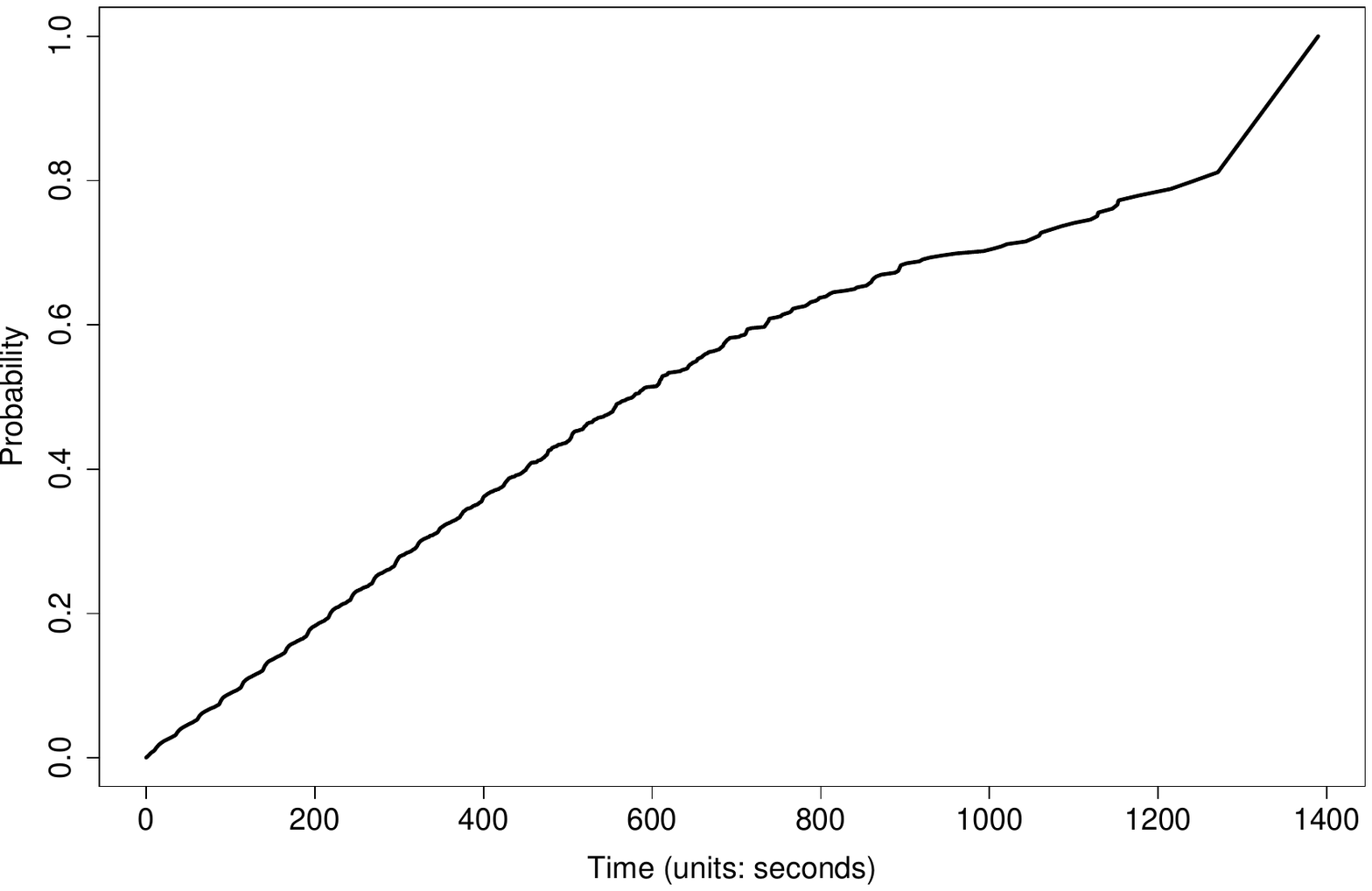}
        \end{subfigure}
        ~ 
        \begin{subfigure}[b]{0.48\textwidth}
                \includegraphics[width=\textwidth]{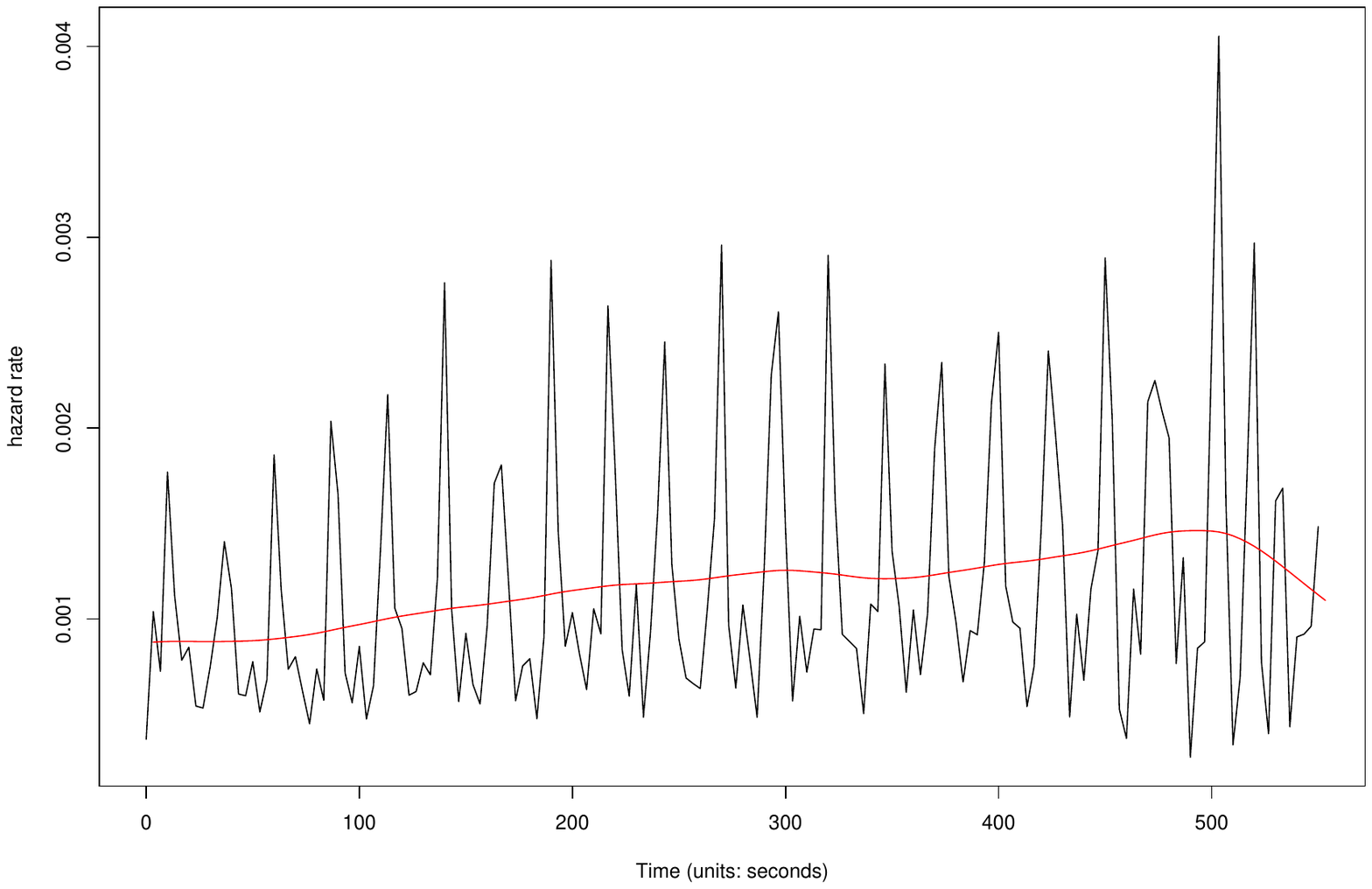}
        \end{subfigure} 
       	 \caption{CDF (left) and hazard rate (right) of the patience}
		\label{fig:patience}
\end{figure}

Previous studies analysing patience distributions include Mandelbaum \& Zeltyn~\cite{mandelbaum2004impact}, Brown et al.~\cite{BrownGMSSZZ} and Roubos \& Jouini~\cite{roubos2013call}, showing constant of slowly decreasing hazard rates. 
That is also what we find, see Figure~\ref{fig:patience} for the empirical distribution function and the hazard rate function (aggregated, in red) of the patience for skill 30175. 
The sudden increase in CDF after 10 minutes waiting is an artefact due to the low number of calls involved. 

The peaks every 25 to 26 seconds in the right plot of Figure~\ref{fig:patience} correspond to the moments when voice messages are announced to all waiting customers. 
Different from the delay announcements studied in Armony et al.~\cite{armony2009impact},  the voice messages in this call center do not give information on the anticipated delay.

\subsection{Breaks}
Shrinkage is a common call center term and is defined as the time an agent gets paid but is not available for handling customer contacts.
Paid breaks are an important part of shrinkage in call centers, together with other activities, such as training, illness, and so forth. 
In this data set, we do not have the original schedule of the agents, thus, we cannot calculate the amount of shrinkage caused by sickness and training. However, we can observe the start times and end times of breaks, as well as whether it is a paid break or an unpaid break (such as a lunch break). In addition, training sessions typically require pre-planning, and illnesses tend to be applied at the beginning of a day. Conversely, breaks, particularly those that are unplanned, are more commonly taken spontaneously throughout the day. 

\begin{figure}[htb]
  \begin{center}
    \includegraphics[width=0.6\textwidth]{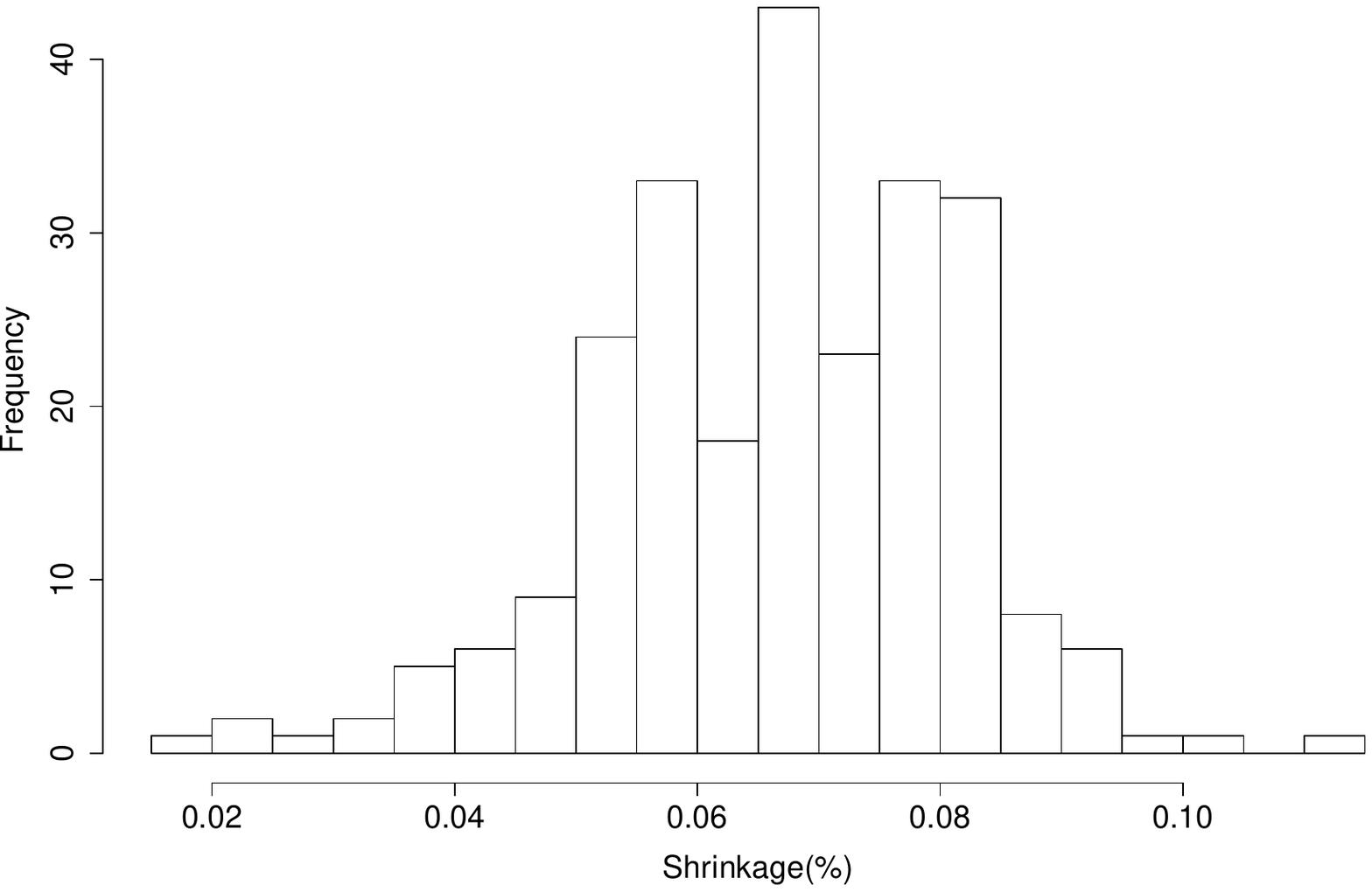}
  \caption{Histogram of shrinkage per agent due to breaks}
  \label{fig:histshrinkage}
  \end{center}
\end{figure}

Shrinkage that is caused by breaks plays a significant part in call center workforce planning. To illustrate the amount of this type of shrinkage, we plot the histogram of the shrinkage per agent in Figure~\ref{fig:histshrinkage}. To generate this plot, we use the following definition:

\begin{align*}
\text{shrinkage} = \frac{\text{time taking breaks}}
{\text{time handling calls, waiting for calls and taking breaks}}.
\end{align*}

The mean shrinkage is 6.7\%, which is quite reasonable. 
In practice, it is uncommon that this type of shrinkage is higher than 10\%. Part of the reason the shrinkage is low is that we do not include agent absenteeism, such as agents being on holiday or sick, or agents doing training. 

\begin{figure}[htb]
  \begin{center}
    \includegraphics[width=0.6\textwidth]{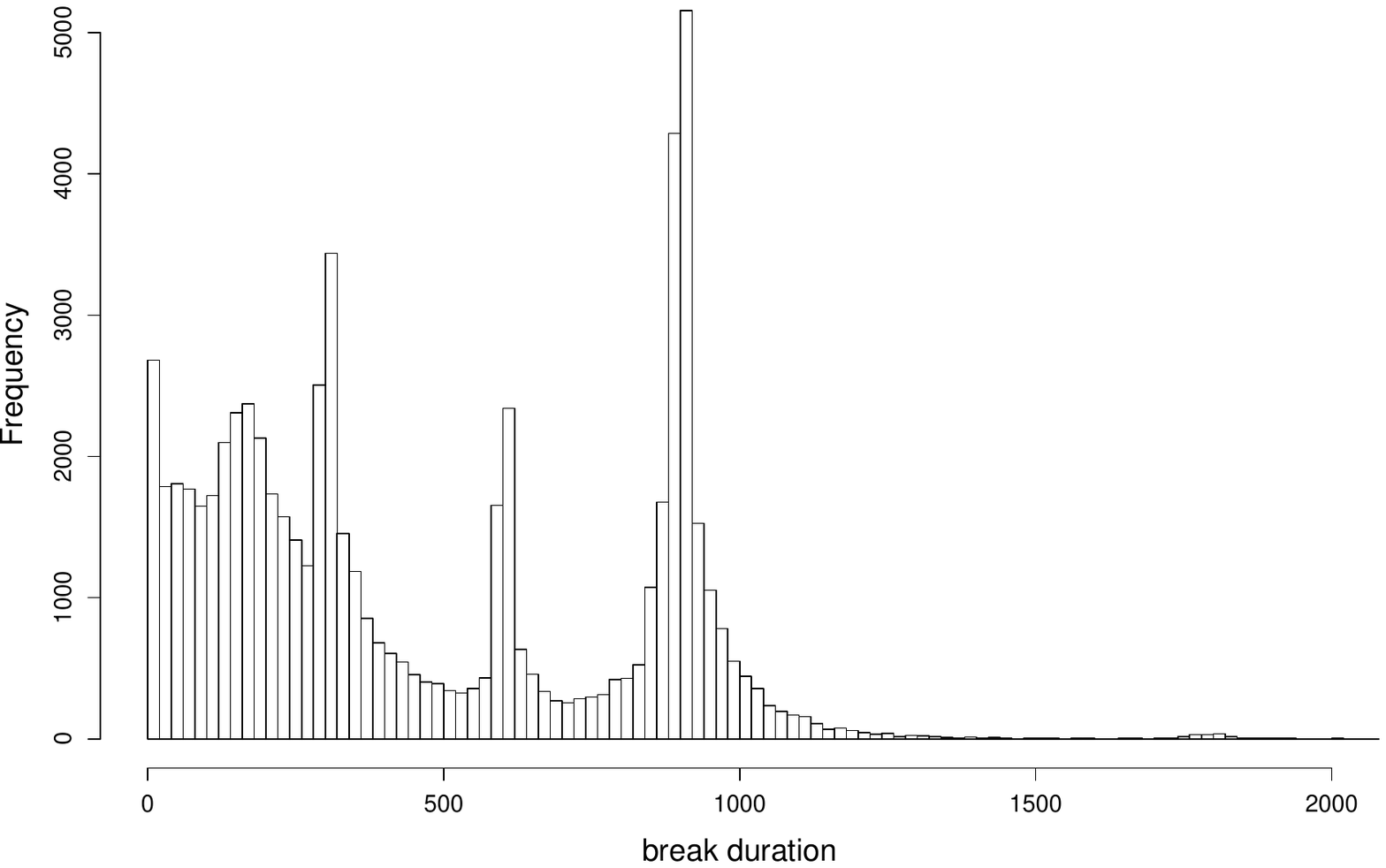}
  \caption{Histogram of break durations (units: seconds).}
  \label{fig:breakdur}
  \end{center}
\end{figure}

The histogram of the break duration is shown in Figure~\ref{fig:breakdur}. One clearly observes peaks at 5, 10 and 15 minutes. The 10 and 15-minute breaks are pre-specified in the shifts, thus, they are in some sense ``plannable". However, they are not completely plannable, because the start times of the breaks and the duration can still have some variation depending on agent preferences and factors such as the business of the call center at that moment. 
For example, some agents prefer taking several 5-minute breaks instead of one 10 or 15-minute full break; another example could be that agents may take less breaks if the call center is currently busy. 
Furthermore, besides these plannable breaks, there are breaks for other purposes, such as agents going to the toilet, having coffee, etc. 
These breaks are usually short and they are ``unplannable". 
An ideal model would differentiate between these types of breaks, then we could study the consequences of ignoring either the ``plannable" or ``unplannable" breaks in SL. 
However, in the data set we have, it is impossible to differentiate. 
For example, suppose that an agent takes a 8-minute break half an hour earlier than the pre-specified break time, then she works for an hour, and then takes a 4-minutes break.
Now it is not clear whether the first and the second breaks are planned or not. 
Therefore, in this paper, we do not make a distinction between these two types of breaks in our models. 

\begin{figure}[htb]
        \centering
	  \begin{subfigure}[b]{0.48\textwidth}
                \includegraphics[width=\textwidth]{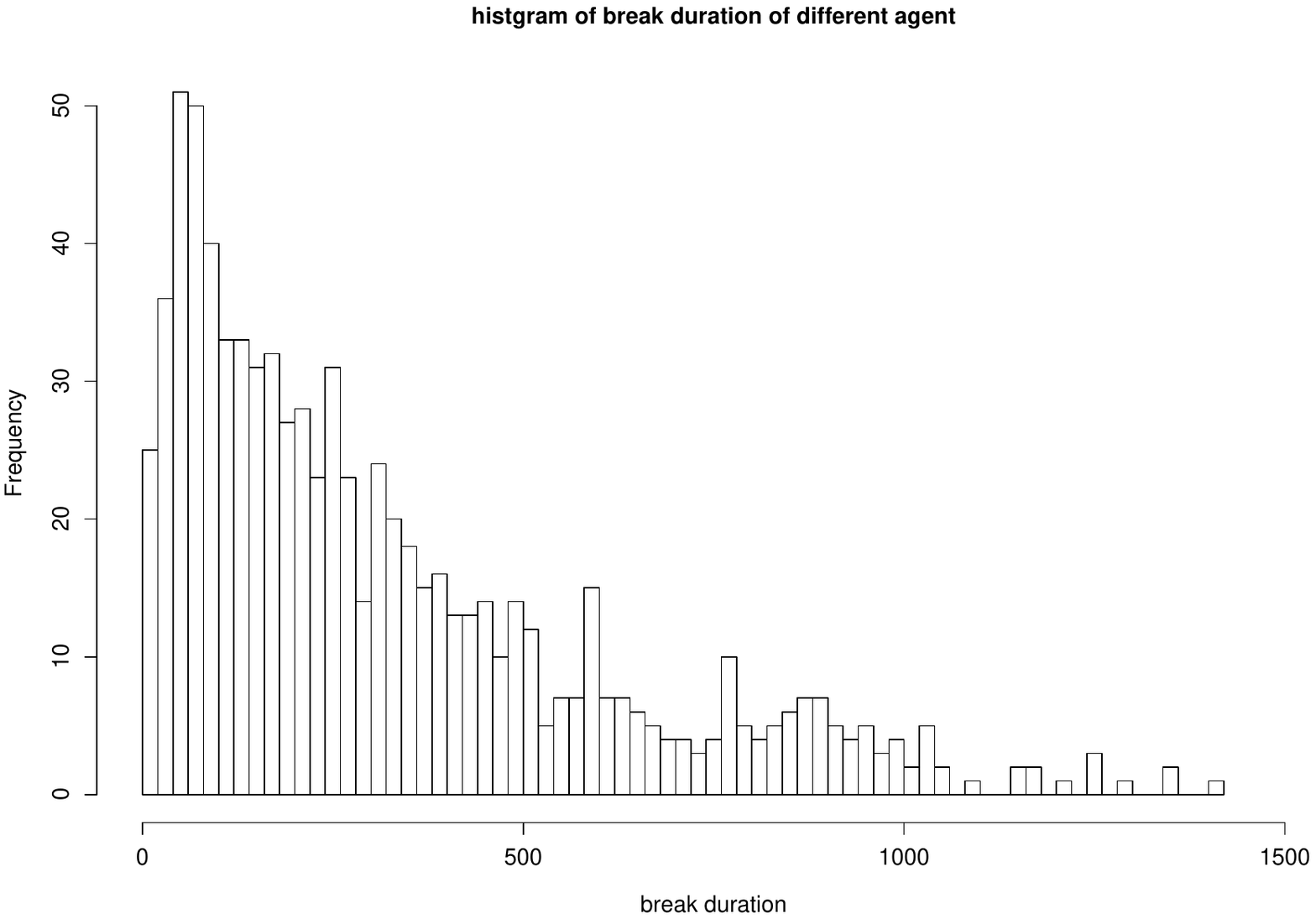}
                \label{fig:Abreak}
        \end{subfigure}
        ~ 
        \begin{subfigure}[b]{0.48\textwidth}
                \includegraphics[width=\textwidth]{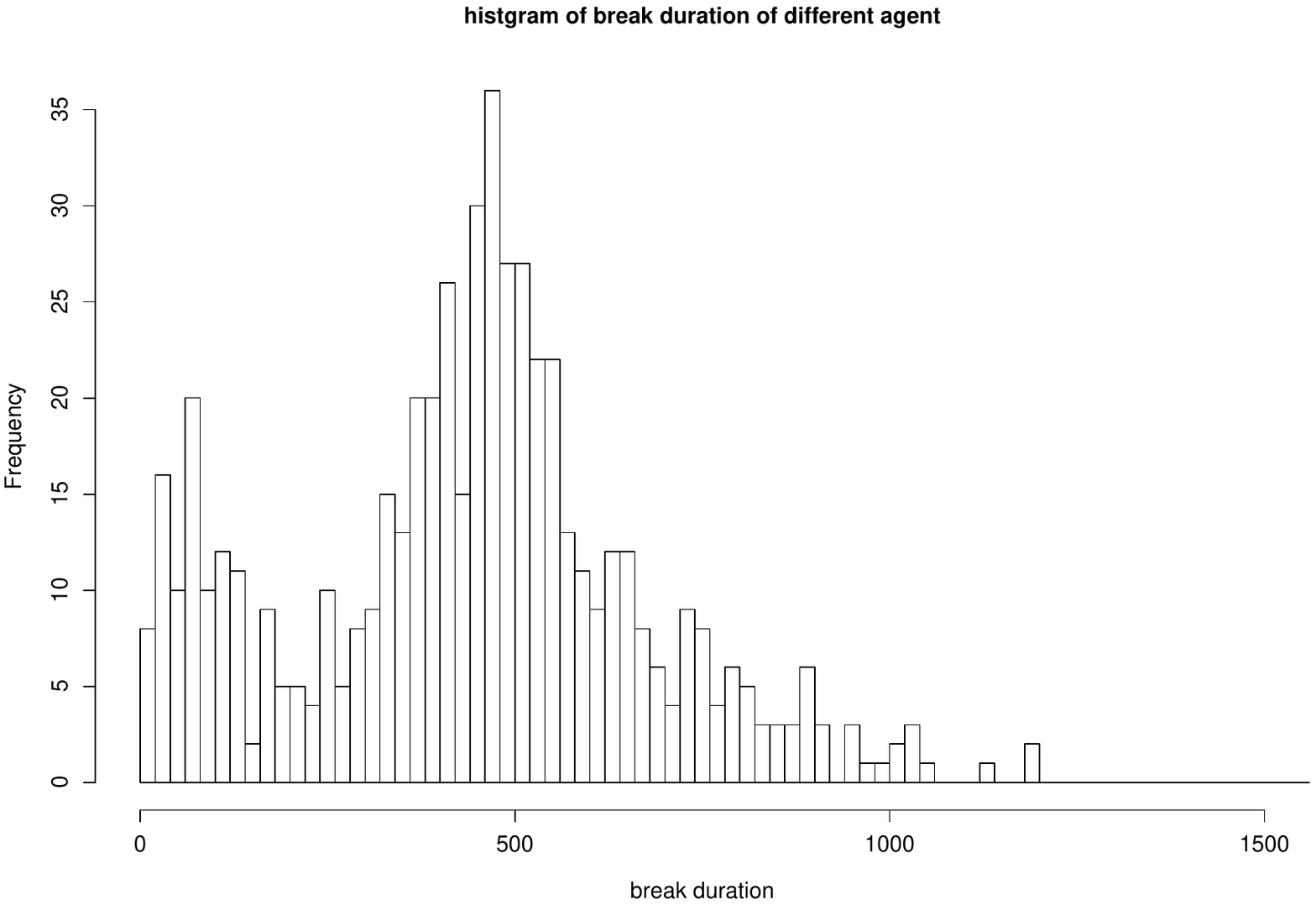}
                \label{fig:Bbreak}
        \end{subfigure} 
       	\\
	  \begin{subfigure}[b]{0.48\textwidth}
                \includegraphics[width=\textwidth]{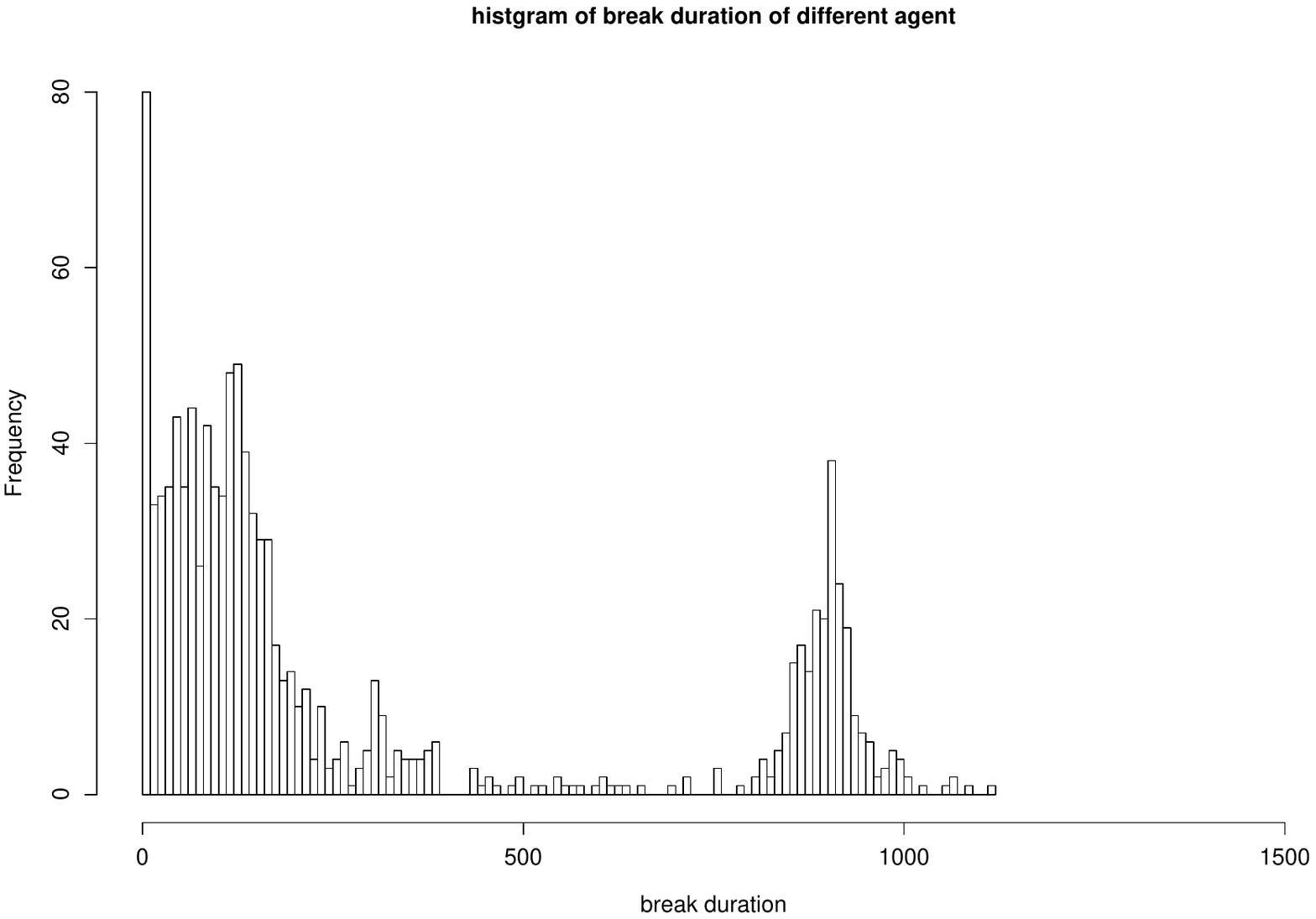}
                \label{fig:Cbreak}
        \end{subfigure}
        ~ 
        \begin{subfigure}[b]{0.48\textwidth}
                \includegraphics[width=\textwidth]{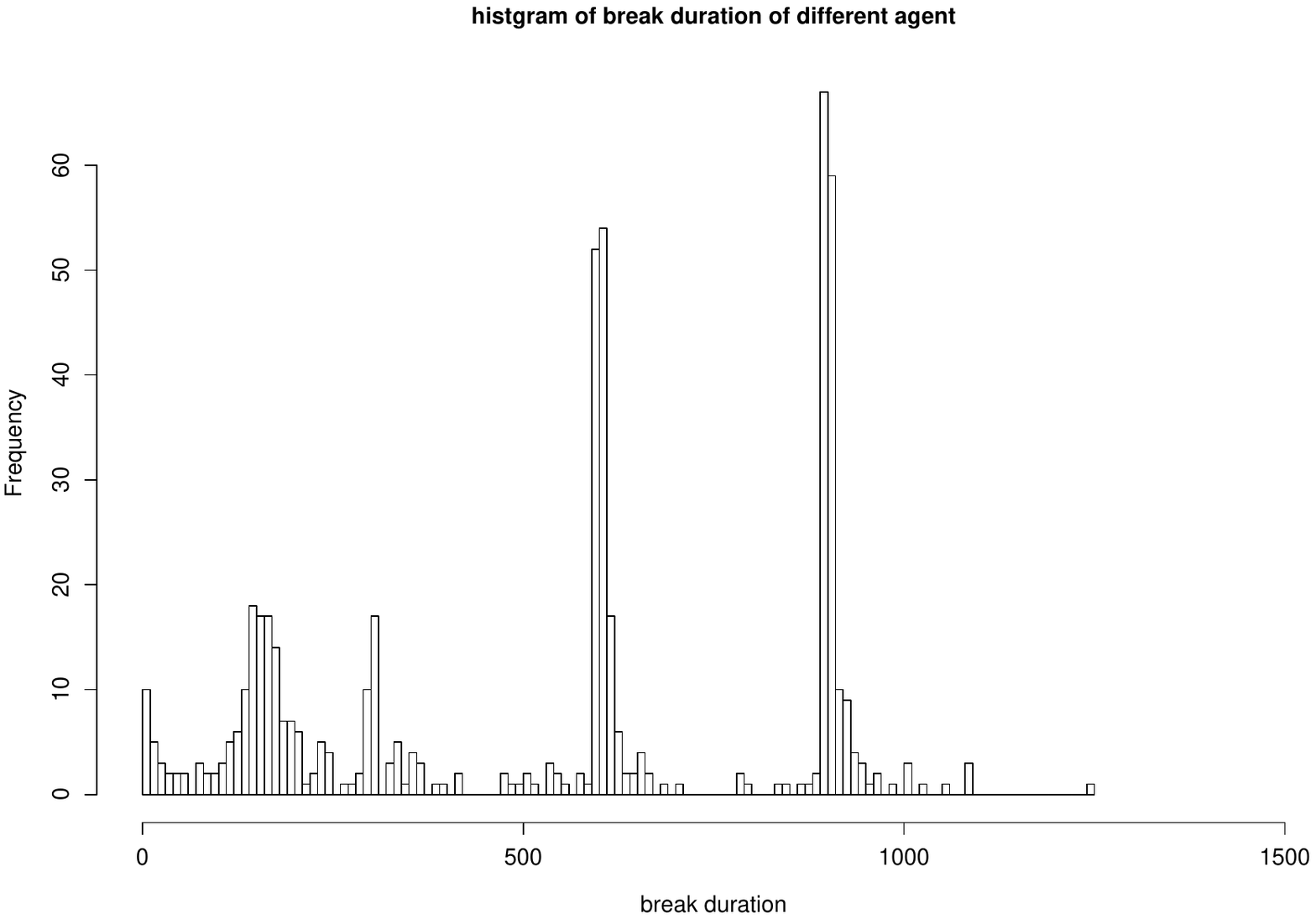}
                \label{fig:Dbreak}
        \end{subfigure} 
       	 \caption{Break duration histograms of some representative agents}
		\label{fig:agentbreak}
\end{figure}

In order to have more insights in break durations, we plotted the break duration histograms of some individual agents in Figure~\ref{fig:agentbreak}. One can conclude from these graphs that different agents have different patterns of break durations. 
For example, the top-left agent prefers short breaks rather than long breaks (we are told by the manager that this agent works at home and is a part-time agent), while the bottom-right agent mostly has 10 and 15-minute breaks with occasionally some small breaks. 


In Section \ref{results} we compare models with and without breaks.

\section{Validating inhomogeneous models} \label{sec:set-up} 

To compare our models to the empirical results we use simulation models.
This is, because we have skill-based operations, a simulation model is the obvious choice.
On top of that, it does not restrict generality, also analytical models based on queueing models such as Erlang A can be simulated.

Before moving to the comparison of the different models we have to address two methodological issues. 
Theory about validating simulation models (see, e.g., Kleijnen \cite{Kleijnen95}) usually assumes a set of i.i.d.\ experiments.
However, our data consists of $n$ realizations of different random variables, because every day in a call center is significantly different from any other.
Therefore we cannot simple compare the averages of both simulations and actuals.
While our model gives the expected performance, the actuals are just realizations. 
To distinguish between the noise of the system and the error of the model we have to estimate the noise of the system and ``subtract" it in the right way from the overall error. 
This noise is substantial, even at the daily level, as shown in Roubos et al.~\cite{RoubosKooleStolletz}.
This is the first issue.

The second is that we want to compare models, apart from the forecasting error we inevitably make.
The only option to eliminate the forecasting error is using the actual values as input for our models. 
But our models assume as input the rate of the NHPP as input, not its realizations.
Thus we have to quantify the error we make by using realizations instead of rates, and again, modify the overall error accordingly.

In this section we propose and motivate a model for quantifying these errors and calculating with them.
We will use this to quantify the model error for our preferred performance measures, which is the mean absolute error in service level.

Let us formulate our model mathematically. 
The r.v.~$\Lambda$ represents the parameters that change from day to day, which are the rate of the non-homogeneous Poisson process and the agents that are scheduled and their shifts. 
The performance, typically the service level obtained during a day, is denoted by $X$. 
Because $X$ depends on $\Lambda$ we write $X(\Lambda)$.  
Note that $X$ is a r.v., even for fixed $\Lambda=\lambda$: its value depends on the realization of the Poisson process, the handling times, times at which agents take breaks, etc.
We can also simulate various models. 
The service level estimation given by the simulation is written as $S(\Lambda)$.
However, the arrival rate is not observed. 
We could replace that part of $\Lambda$ by a forecasts, but they are usually quite bad.
Instead, we use use the actual instead of the rates. 
$\Lambda$ in which the rates are replaced by the random realizations of the arrivals is written as $A(\Lambda)$, the corresponding simulation $S(A(\Lambda))$.

With $\E_\bullet$ we indicate the expectation with respect to the corresponding r.v. 
For example, $\E_SS(A(\Lambda))$ is the expected simulated performance of a random day for a random realization of the rates; $\E_\Lambda X(\Lambda)$ is the random performance ``averaged" over the days.
Note that we can interchange expectations, e.g., $\E_X\E_\Lambda X(\Lambda)=\E_\Lambda\E_XX(\Lambda)$.

We are interested in estimating $\E_\Lambda|\E_XX(\Lambda)-\E_SS(\Lambda)|$, which corresponds to the mean absolute error (MAE) of the service level. However, we measure $X(\Lambda)$ and $\E_SS(A(\Lambda))$, the latter by averaging over a sufficiently high number of simulations. 
We will show how to get an estimate of the MAE based on the actuals and simulations.

We have
\[\E_SS(A(\Lambda))-X(\Lambda)=\E_SS(A(\Lambda))-\E_SS(\Lambda)+\E_SS(\Lambda)-\E_XX(\Lambda)+\E_XX(\Lambda)-X(\Lambda).\]
The l.h.s.~is observed. In the r.h.s.~there are 3 differences, respectively:\\
-  $\E_SS(A(\Lambda))-\E_SS(\Lambda)$, which we call the {\em cheating factor}, because you cannot use the actuals in the simulation as at the moment of planning they are not known;\\
- $\E_SS(\Lambda)-\E_XX(\Lambda)$, the {\em model error}, which we would like to estimate;\\
- $\E_XX(\Lambda)-X(\Lambda)$, the {\em system noise}. 

Both the cheating factor and the systems noise cannot be approximated directly. Let us start with the system noise. 
$\E_XX(\Lambda)$ is not available because we have only 1 realization for each possible value of $\Lambda$, i.e., every day.
But we can approximate it using simulations, as we expect $\E_XX(\Lambda)-X(\Lambda)\approx\E_SS(\Lambda)-S(\Lambda)$.
Using the latter formula we obtain
\[\E_SS(A(\Lambda))-X(\Lambda)\approx\E_SS(A(\Lambda))-S(\Lambda)+\E_SS(\Lambda)-\E_XX(\Lambda).\]

$\E_SS(A(\Lambda))-S(\Lambda)$ can be seen as a combination of the cheating factor and the system noise. 
We cannot obtain it directly because $S(\Lambda)$ is a simulation with an unknown arrival rate. 
Instead, we use $A(\Lambda)$ as rates and sample again, leading to $A^2(\Lambda)$. 
Using simulations on comparable rates show that $\E_SS(A^2(\Lambda))-S(A(\Lambda))\approx\E_SS(A(\Lambda))-S(\Lambda)$. 

This leads to 
\begin{equation}
\E_SS(A(\Lambda))-X(\Lambda)\approx\E_SS(A^2(\Lambda))-S(A(\Lambda))+\E_SS(\Lambda)-\E_XX(\Lambda).
\label{error-eq}\end{equation}

\begin{figure}[htb]
  \begin{center}
    \includegraphics[width=0.9\textwidth]{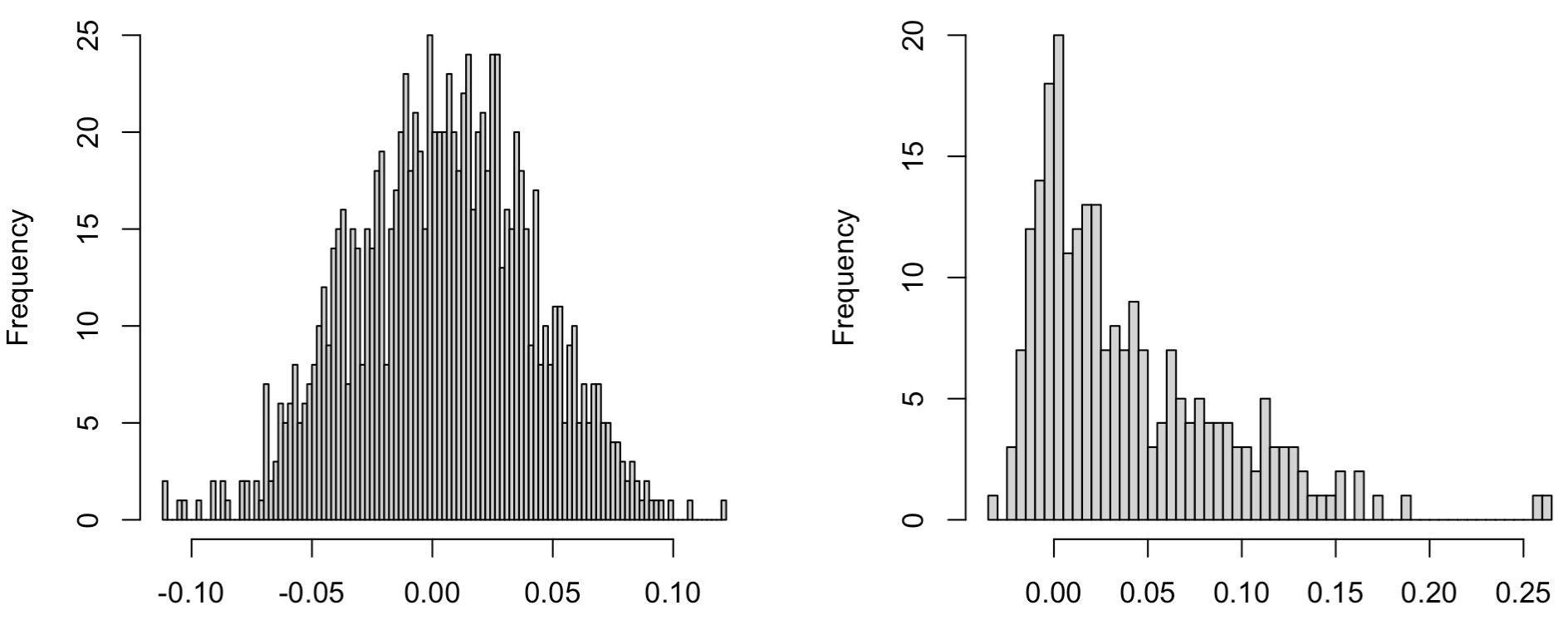}
  \caption{Histograms of $\E_SS(A^2(\Lambda))-S(A(\Lambda))$ and $\E_SS(A(\Lambda))-X(\Lambda)$.}
  \label{fig:cheating}
  \end{center}
\end{figure}

This expression can be used to approximate the distribution of the model error.
It is interesting to note that $\E_SS(A^2(\Lambda))$ and $S(A(\Lambda))$ are positively correlated: if $A(\Lambda)>\Lambda$ then we expect $S(A(\Lambda))$ to be low. 
At the same time $\E_SS(A^2(\Lambda))$ will be low because it uses the high value of $A^2(\Lambda)$. 
This means that the noise of the simulation is partially canceled by the cheating leading to, in our case, a relative small joint effect. 
As we can see from the left figure of Figure \ref{fig:cheating} $\E_SS(A^2(\Lambda))-S(A(\Lambda))$ is approximately normally distributed with a mean equal to 0.4\% and a standard deviation of 3.7\%.
Its MAE is $\E_{\Lambda,A}\big|\E_SS(A^2(\Lambda))-S(A(\Lambda))\big|=3\%$.

Our goal is to compute the MAE of $\E_SS(\Lambda)-\E_XX(\Lambda)$, $\E_\Lambda\big|\E_SS(\Lambda)-\E_XX(\Lambda)\big|$. 
We cannot simply subtract 3\% from $\E_\Lambda\big|\E_SS(A(\Lambda))-X(\Lambda)\big|$. 
To get a better understanding we start with computing the first two moments of the model error.
Our simulations show $\E_{\Lambda,A}\big(\E_SS(A(\Lambda))-X(\Lambda)\big)=3.9\%$ and $\sigma_{\Lambda,A}\big(\E_SS(A(\Lambda))-X(\Lambda)\big)=5\%$. 

The first moment of the model error follows directly by taking expectations in Equation~(\ref{error-eq}):
\[\mu:=\E_{\Lambda}\big(\E_SS(\Lambda)-\E_XX(\Lambda)\big)=3.5\%.\]
This value is of interest in itself: it tells us to which extend the model is {\em biased}, to which extent there is a systematic error.
But even if $\mu$ is small, the errors can be big but fluctuating, sometimes positive, sometimes negative. 
That is why we defined the performance measure as $\E_\Lambda\big|\E_SS(\Lambda)-\E_XX(\Lambda)\big|$.

Next we compute the standard deviation of the model error.
As we expect $\E_SS(A^2(\Lambda))-S(A(\Lambda))$ to be independent of the model error, we have
\[\sigma^2:=\sigma^2_{\Lambda,A}\big(\E_SS(\Lambda)-\E_XX(\Lambda)\big)\approx\]
\[\sigma^2_{\Lambda,A}\big(\E_SS(A(\Lambda))-X(\Lambda)\big)-
\sigma^2_{\Lambda,A}\big(\E_SS(A^2(\Lambda))-S(A(\Lambda))\big).\]
In our case $\sigma=\sqrt(0.05^2-0.037^2)=3.4\%$.
We conclude that the first moments of the model error and will also be quite similar to the measured error.

We can go a step further if we assume the measurements to be normally distributed. For the simulated noise/cheating factor this is the case, for the measured error this is a rough approximation, as can be seen from the right plot of Figure \ref{fig:cheating}.
For $\E_SS(\Lambda)-\E_XX(\Lambda)\sim{\rm N}(0.035,0.034)$, using a straightforward calculation, we find 
\[\E_\Lambda\big|\E_SS(\Lambda)-\E_XX(\Lambda)\big|\approx{2\sigma\over\sqrt{2\pi}}{\rm e}^{-{1\over2}({\mu\over\sigma})^2}+\mu\Big(2\Phi\Big({\mu\over\sigma}\Big)-1\Big)=4.0\%,\]
with $\Phi$ the standard normal distribution function.
Note that the MAE without the correction is $\E_{\Lambda,A}\big|\E_SS(A(\Lambda))-X(\Lambda)\big|=4.5\%$.
We conclude that the added precision given by the correction is small compared to the measured error, around 10\%.
For this reason we will report on the measured errors from now on.

The weak point in this analysis is the normal assumption for the measured errors.
Using characteristic functions of the empirical distributions we could numerically derive the distribution of the model error and then sample the MAE.
Although interesting, we believe this is beyond the scope of this paper.

The results of this section are derived for the ``HT \& Patience Model", see the next section for its meaning and a comparison with other models. These models give similar results. Note that even the best model will give a non-zero measured error. 
The lower bound to the measured error is approximated by $\E_SS(A^2(\Lambda))-S(A(\Lambda))$.
Its MAE is 3\%.

We will also use other performance measures, such as the MAE of the mean waiting time and the fraction of abandoned calls. They also give comparable results.

\section{Validation results} \label{results}

In this section we compare nine models which differ in the way the properties of Section~\ref{sec:analysis} are modeled. 
By comparing them we can gain a better understanding of the impact of different assumptions. 
These insights can be beneficial for practitioners in evaluating performance in real-world scenarios.

We divide the opening hours of the call center into 24 intervals of 30 minutes. The number of arrived calls and the number of agents working are then calculated per interval and used as input in simulation.
Both the service level (SL) and abandonment rate are calculated using the total offered calls as the denominator. We set the acceptable waiting time (also called {\em time-to-answer}) to 60 seconds.
After removing the regular closed days, holidays and some days that have call/activity log errors (due to technical issues), we have in total 237 days. Each day, we repeat the simulation 1000 times and compare the 1000 simulation outcomes with the actual derived from the log file.

The assumptions we considered are as follows:

\begin{description}
\item[Arrivals:] Arrivals set to {\bf ``Identical"} means that the arrival processes are identical in simulation and in real data, i.e., if there is an arrival at time $t$ in real data, we schedule an arrival at the exact same moment in the simulation. Arrival is set to {\bf ``IPP"} stands for in-homogeneous Poisson process with piecewise constant rate, which means that if there were $a$ arrivals within a certain interval in the data, then we schedule $a'$ arrivals within that interval in each simulation run, with $a'\sim{\rm Poisson}(a)$.
\item[HT, AHT per day:] HT set to {\bf ``Empirical"} means that we use the empirical handling time distribution (per skill) in the simulation. The empirical distribution is derived from all calls of the whole year if {\bf AHT per day = ``No"}, or it is day-specific considering only the calls handled on that day if {\bf AHT per day = ``Yes"}. HT set to {\bf``Exp"} means that we assume that the HT follows an exponential distribution (as assumed in all Erlang models) with its mean being the average of the actual HTs of calls (per skill) over the day if {\bf AHT per day = ``Yes"}, or fitted using the method explained in Section \ref{HT-sec} if {\bf AHT per day = ``Fit"}.
\item [Wrap-up:] Wrap-up set to {\bf ``Yes"} means that in the simulations we add the average wrap-up time to the HT. Wrap-up is set to {\bf ``No"} means we do not consider wrap-up times.
\item [Patience:] Patience set to {\bf ``Empirical"}, means that, for each customer in the simulation, we generate a random patience time from the empirical CDF estimated by the Kaplan-Meier estimator. Patience is set to  {\bf ``Exp"} means that we assume the patience follows an exponential distribution, with its mean being the empirical mean of the patience from the data, estimated via the Kaplan-Meier estimator.
\item [(paid) Breaks:] Breaks set to {\bf ``Yes"} means that if an agent takes a break, then we subtract the proportional staffing levels from the total staffing levels of this interval. 
Breaks set to {\bf ``No"} means we do not consider paid break shrinkage in the model. For example, in a certain interval, the 6 agents are working (either waiting for a call, answering a call or wrapping up). Therefore, the total working time should be 180 minutes, but two agents had breaks during this interval and each break lasted 10 minutes. If Breaks is set to ``Yes", then we assume there were in total (180 - 20)/30 = 5.3 agents working in this interval, and we round it to 5 in the simulation. If Breaks is set to ``No", then we ignore the breaks and assume in the simulations that there were in total 180/30 = 6 agents working in this interval.
\end{description}

Note that besides working on these eight skills that we selected and taking breaks, there are other agents' activities, such as working on calls of other skills, making outbound calls, having consultations with managers or other senior agents, etc. The amount of time that spends on these activities is very little compared to the time spent on breaks.
We exclude these durations in a similar way as we remove the agent breaks.

The models with their assumptions are shown in Table~\ref{table:models}.

\begin{table}[htb]
  \begin{center}
\resizebox{\linewidth}{!}{%
  \begin{tabular}{@{}lcccccc@{}}
	\toprule
 	 & Arrival & HT & AHT per day& Wrap-up & Patience & Breaks \\
   	\midrule
Empirical Model & Identical & Empirical & Yes & Yes & Empirical & Yes \\
Arrival Model & IPP & Empirical & Yes & Yes & Empirical & Yes \\
Daily HT Model & IPP& Exp & Yes & Yes & Empirical & Yes \\
Fitted HT Model & IPP & Exp & Fit & Yes & Empirical & Yes \\
Yearly HT Model & IPP& Empirical & No & Yes & Empirical & Yes \\
Patience Model & IPP& Empirical & Yes & Yes & Exp & Yes \\
HT \& Patience Model & IPP& Exp & Yes & Yes & Exp & Yes \\
Breaks Model & IPP& Empirical & Yes & Yes & Empirical & No \\
Wrap-up Model & IPP& Empirical & Yes & No & Empirical & Yes \\
    \bottomrule
  \end{tabular}}
  \caption{Models.}
  \label{table:models}
  \end{center}
\end{table}

\begin{description}
\item[Empirical Model:] In this model, our objective is to make assumptions that are as close to reality as possible. In other words, we aim to minimize model errors. To do this, we replicate the exact arrival moments in the simulation model and use the empirical distribution of HTs per day and the empirical distribution of patience. 
The average wrap-up time is added to HTs. 
We also consider breaks by calculating the proportional staffing levels. 
This model serves as a benchmark for validation, presumably setting the lower bound of the MAE.

\item[Arrival Model:] The only difference between this model and the Empirical Model is the arrival process.
In this model the follow the IPP with the rate per interval equal to the actual arrivals in that interval. 
The impact of the assumption of ‘IPP’ arrivals on the performance metrics prediction can be then analysed by comparing these two models. As one can see that from Table \ref{table:per1}, the small difference between the MAEs of the two models suggests that whether the arrival process is ‘Empirical’ or ‘IPP’ does not have a strong influence on the accuracy in predicting performance measures SL, Ab and ASA. Therefore, we update our benchmark to be the Arrival Model for simplicity.

\item[Daily HT Model:] The only difference in the assumptions between Arrival Model and this Daily HT Model is how we handle the HT assumption. It is designed to assess the influence of the exponential assumption on the HTs. 

\item[Fitted HT Model:] The only difference in the assumptions between this model and the Daily HT Model is how the daily AHT is calculated. In the Daily HT Model, the daily AHT per skill is just the average of all calls from that skill handled on that day. 

\item[Yearly HT Model:] This model is designed to assess the influence of disregarding the day effect of the HTs. The only difference between this model and the Arrival Model is that the AHT per day is set to No. In other words, the empirical HT distribution is derived using the calls of the whole year. 

\item[Patience Model:] This model differs from the Arrival Model only in the patience assumption. 
It is designed to assess the influence of the exponential assumption of the patience.
Unlike the Arrival Model, where the patience follows directly the empirical distribution derived from applying the Kaplan-Meier method, in this model, patience follows the exponential distribution with the mean equal to the mean of the empirical distribution. 

\item[HT \& Patience Model:] This model is designed to reflect the assumptions made by the Erlang models. HT and patience are both exponentially distributed. Arrivals follows Poisson distribution. Moreover, by comparing this model with the previous Patience Model, HT and Patience we can discuss whether the exponential assumption of the patience distribution has a stronger influence on the accuracy of the models comparing to the exponential assumption of the HT.

\item[Breaks Model:] This model is introduced to assess the influence of not taking into account small paid breaks when predicting the performance. The effect of agent breaks in the accuracy can be concluded by comparing to the Arrival Model. 

\item[Wrap-up Model:] This model differs from Arrival Model only in wrap-up time assumption. It is designed to understand whether adding the short wrap-up time to HT in the WFM models will lead to a more accurate prediction. 

\end{description}

Besides the MAE, we also determine for each model the percentage $\text{I}_\alpha$ of the actuals that are within the $\alpha$-confidence interval of the simulation outcomes, for $\alpha = 95\%$. 
Furthermore, we compute the fraction of actuals that is higher than the median of the simulation outcomes.
This gives an idea to what extend the outcomes are biased.

The MAE of SL, Abandonments (Ab) and Average waiting time (called {\em Average speed of answer} (ASA) in call centers) are shown in Table~\ref{table:per1}. 
The unit of $\text{MAE}_{ASA}$ is seconds. 
The percentage of the actuals that are within the 95\% confidence interval of the simulation outcomes is also displayed.
Table~\ref{table:variab} lists the variability of the simulation measurements.
The percentage of actuals that are above the median of the predicted SL, Ab and ASA in each model is shown in Table \ref{table:actual}.

%
 
\begin{table}[htb]
\begin{center}
\resizebox{\linewidth}{!}{%
  \begin{tabular}{@{}lcccccc@{}}
	\hline
 	& \multicolumn{1}{c}{$\text{MAE}_{SL}$} &  \multicolumn{1}{c}{$\text{MAE}_{Ab}$}  &  \multicolumn{1}{c}{$\text{MAE}_{ASA}$} & $\text{I}_{\alpha, \text{SL}}$ & $\text{I}_{\alpha, \text{Ab}}$& $\text{I}_{\alpha, \text{ASA}}$ \\
   	\hline
Empirical Model & $ 3.01\%$ & $0.83 \%$ & $7.57$ &$66.0\%$& $37.9\%$ &$51.0\%$ \\
Arrival Model & $3.03\%$ & $0.77 \%$ & $6.71$ &$86.4\%$& $78.9\%$ &$76.7\%$ \\
Daily HT Model & $2.93\%$ & $0.67  \%$ & $6.14$  &$92.4\%$& $88.0\%$ &$89.0\%$ \\
Fitted HT Model & $4.10\%$ & $1.03  \%$ & $9.19$  &$81.8\%$& $74.6\%$ &$76.7\%$ \\
Yearly HT Model & $5.70\%$ & $1.54 \%$ & $12.49$  &$64.0\%$& $55.0\%$ &$55.9\%$ \\
Patience Model & $4.88\%$ & $0.65  \%$ & $12.35$&$62.8\%$& $83.9\%$ &$46.4\%$ \\
HT \& Patience Model & $4.54 \%$ & $0.56  \%$ & $11.83$  &$73.4\%$& $90.2\%$ &$57.3\%$ \\
Breaks Model& $8.50\%$ & $2.00  \%$ & $17.34$  &$31.0\%$& $15.6\%$ &$16.8\%$ \\
Wrap-up Model &  $4.23\%$ & $1.10  \%$ & $9.40$  &$72.5\%$& $48.9\%$ &$57.8\%$ \\
  \hline
  \end{tabular}}
  \caption{Performance measures of models.}
  \label{table:per1}
  \end{center}
 \end{table}
 
\begin{table}[htb]
  \begin{center}
  \resizebox{\linewidth}{!}{%
  \begin{tabular}{@{}lccc@{}}
	\toprule
 	& Variability of SL  & Variability of Ab  & Variability of ASA\\
   	\midrule
Empirical Model &$1.67\%$ & $0.27 \%$  & $2.37$ \\
Arrival Model &$2.47\%$ & $0.45 \%$  & $3.76 $\\
Daily HT Model&$2.73\%$ & $0.52 \%$  & $4.32$\\
Fitted HT Model&$2.02\%$ & $0.43 \%$  & $4.26$\\
Yearly HT Model &$2.48\%$ & $0.41 \%$ & $3.63$\\
Patience Model &$4.01\%$ & $0.59\%$  & $2.94$ \\
HT \& Patience Model &$0.24\%$ & $0.50 \%$ &$3.43$\\
Breaks Model&$1.90\%$ & $0.30 \%$  & $2.82$\\
Wrap-up Model &$2.30\%$ & $0.39 \%$  & $2.54$\\
    \bottomrule
  \end{tabular}}
  \caption{Variability of performance measures}
  \label{table:variab}
  \end{center}
\end{table}

\begin{table}[htb]
  \begin{center}
  \resizebox{\linewidth}{!}{%
  \begin{tabular}{@{}lccc@{}}
	\toprule
 	& $\p(\text{SL} > Q_{0.5, \text{SL}})$  & $\p(\text{Ab} > Q_{0.5, \text{Ab}})$  & $\p(\text{ASA} > Q_{0.5, \text{ASA}})$\\
   	\midrule
Empirical Model &$29.1\%$ & $83.1\%$  & $84.4\%$ \\
Arrival Model & $ 24.5\%$  &$ 84.1\%$ & $85.0\%$\\
Daily HT Model & $33.8\%$  &$78.9\%$ & $78.1\%$ \\
Fitted HT Model & $34.4\%$  &$72.6\%$ & $75.9\%$ \\
Yearly HT Model & $35.0\%$  &$69.6\%$ & $71.3\%$\\
Patience Model& $ 12.7\%$ &$75.9\%$ & $94.5\%$ \\
HT \& Patience Model & $18.1\% $  &$ 70.9\%$ & $90.7\%$ \\
Breaks Model & $21.1\%$  &$98.7\%$ & $98.7\%$\\
Wrap-up Model & $13.5\%$  &$94.9\%$ &  $95.4\%$ \\
    \bottomrule
  \end{tabular}}
  \caption{Percentage of actuals above the median.}
  \label{table:actual}
  \end{center}
\end{table}

We describe now the observations and insights we draw by comparing these models. 
The Empirical Model, Arrival Model and Daily HT Model are comparable, with MAE about 3\% in SL, 0.7\% in Ab, and 6.8 seconds in ASA, despite the fact that we made certain simplifications, such as rounding the number of agents in each interval, and no redials after abandonment (see Ding et al.~\cite{DingKooleMei-truedemand}). 
This means that whether the arrival process is ``Identical" or ``IPP" does not have a strong influence on the performances of the models. Thus, if the forecast for arrival rate is exact, assuming in-homogeneous Poisson arrival processes will not degrade the model.
Furthermore, the exponential assumption of HT does not have a strong influence on the performance of the models either, as long as the AHT is the same. This is conform results by Whitt~\cite{Whitt-ms05-engineering}.

Interestingly, the Daily HT Model with exponential handling times performs slightly better than the Arrival Model with empirical handling times. 
The reason is as follows.
All models we consider are slightly ``optimistic" comparing to reality, in the sense that they predict higher SL, lower Ab and ASA compared to the actuals. 
This can be concluded from the numbers in the first column of Table \ref{table:actual} which are all below 50\%, and the next two columns are all above 50\%. 
This is especially true in the days when the actual SL is low and the actual number of handled calls is also low.
Using the exponential assumption will lead to a more conservative model, since exponential HTs have a higher variance than the empirical distribution, and therefore the simulated performance is more close to the actual performance.

Comparing the Fitted HT Model and the Daily HT Model, one can easily see the extra performance error introduced by the fitted error of AHT. 
However, the Fitted HT Model still gives smaller MAE than the Yearly HT Model. 
This means that the day effect in the handling times should not be disregarded.

Now let us look at the Patience Model.
In the first place, the exponential assumption of the patience distribution has a stronger influence on the accuracy of the models compared
 to the exponential assumption of the HT. This can be concluded by comparing the Arrival Model, the Daily HT Model, the Patience Model and HT \& Patience Model. 
 This finding is in line with Whitt~\cite{Whitt-ms05-engineering}.
There it is theoretically shows that the behavior of some queueing models (such as the $M/GI/s/r+GI$) is primarily affected by the HT distribution through its mean while it is primarily affected by the patience distribution by its hazard function near the origin, and not by its mean or tail behavior.

\begin{figure}[htb]
  \begin{center}
    \includegraphics[width=\textwidth]{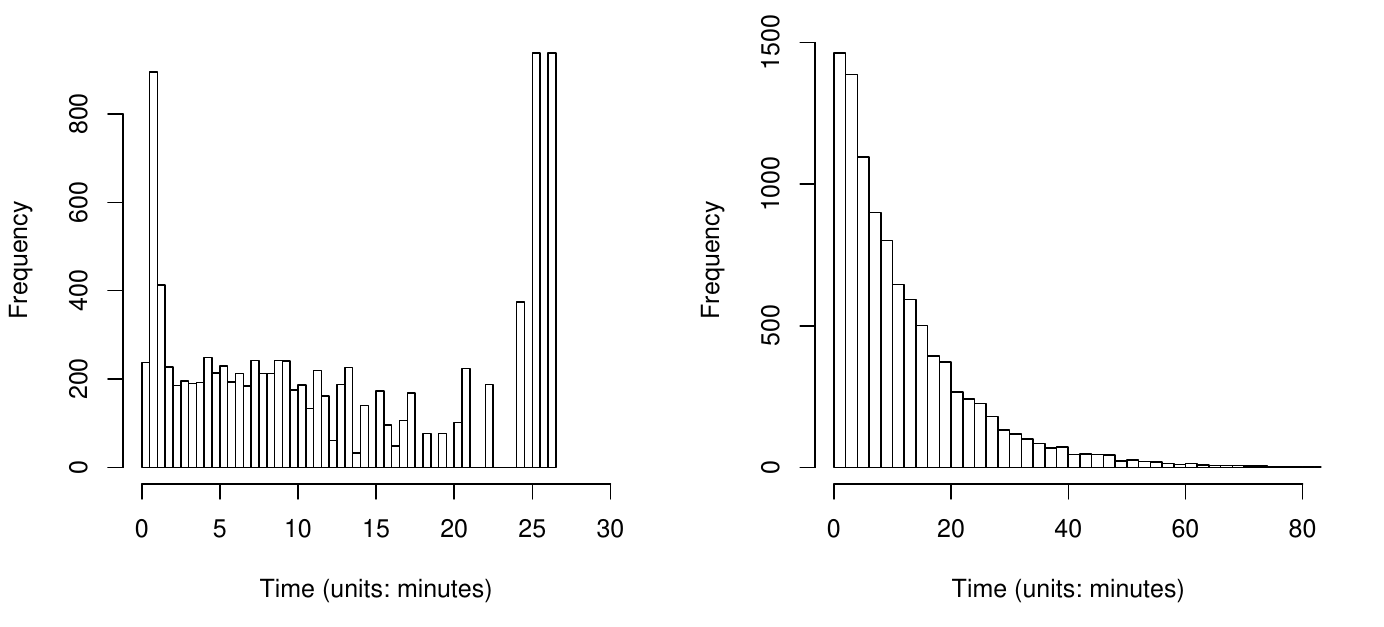}
  \caption{Histogram of the patience samples from the empirical distribution (left) and the exponential distribution (right).}
  \label{fig:twopatience}
  \end{center}
\end{figure}

In the second place, we observe that the variability of the Patience Model (especially the SL) is almost doubled compared to the Arrival Model in Table \ref{table:variab}. 
This is related to the fact that the variance of the exponential distributed patience is much higher (220\%) than the variance of the empirical patience distribution. 
Moreover, the two distributions have quite different structures, see Fig \ref{fig:twopatience}. 
The probability of a short patience in the exponential distribution is higher than the empirical distribution, therefore, more abandonments will occur which is more close to the actual performance. This can be seen from the second column of Table \ref{table:actual}: $\p(\text{Ab} > Q_{0.5, \text{Ab}})$ of the Patience Model is smaller than the Arrival Model. However, more abandonments can result in a higher SL, which leads to a higher MAE of SL: see also $\p(\text{Ab} > Q_{0.5, \text{Ab}})$.

Among all models, the Breaks Model gives the highest error, which implies that taking agent breaks into account has a drastic influence on the accuracy of the models. 
As one can see, by comparing the Arrival Model and the Break Model, the error is nearly more than doubled for the SL and even more than doubled for the Ab and ASA. 
This is not surprising since considering breaks or not is directly related to the staffing levels used in the simulation. It implies that a model that can capture agent patterns of paid breaks will help to improve the accuracy of performance prediction.

Finally, although the wrap-up times are in general very short in a call center, ignoring them could still lead to unnecessary inaccuracy. As one can see from the comparison between the Wrap-up Model and the Arrival model: without considering wrap-up times, the error obviously increases.

\section{Conclusion} \label{sec:conclusion}

In this paper, we validated several WFM staffing models for multi-skill call centers. The validation is done by comparing predictions based on simulations in SL, Ab and ASA to the actual values. 
The comparison results as well as the empirical analysis results show that several properties in call centers, such as agent breaks, wrap-up times, AHT variability are essential for making precise predictions. 
Furthermore, we also verify some of the commonly used key assumptions and simplifications in call center models, such as rounding the number of agents per interval, assuming in-homogeneous Poisson arrival processes, and the exponential assumption for the handling times and customer patience. 
We quantify the influence of these assumptions in terms of prediction errors in SL, Ab, and ASA. 
It turns out that the in-homogeneous Poisson arrival processes assumption and exponential assumption of the HT do not have a significant influence on the accuracy of the model, while the exponential assumption of the patience does. 
Last but not the least, we empirically show that the AHT of each agent differs, and the AHT of new agents decrease as they learn over time. We then develop a model to fit and predict the AHT of each day for each agent. 
These two effects partially explain the variability in AHT of each day, and a staffing model with fitted AHT leads to large improvements compared to a model that ignores such variability.

Several topics can be interesting for further research: 
1) The data we have is from a multi-skill call center. It would be interesting to analyze single-skill call center data, and validate single-skill staffing models, such as the Erlang C and Erlang A models. 
2) Redial (re-attempt after abandonment) and reconnect (re-attempt after connections) behavior is not included in the models, because we cannot identify customers in the data. 
One extension would be to study models with redials and reconnects. 
3) Shrinkage (non-productive time such as meetings and paid breaks) has a big impact on call center performance. It would be interesting to study deeper into all types of shrinkages and the shrinkage patterns of different agents.
\medbreak

\par{\bf Acknowledgements} We thank VANAD Laboratories for supplying the data and Rob van der Mei and Raik Stolletz for useful discussions.

\bibliographystyle{plain}
\bibliography{refs} 

\end{document}